\documentclass{aa}
\usepackage{graphicx}
\usepackage[varg]{txfonts}
\usepackage{natbib}
\usepackage{upgreek}
\usepackage{multirow}
\bibpunct{(}{)}{;}{a}{}{,}

\begin{document}
   \title{Modelling depletion by re-accretion of gas from a dusty disc in post-AGB stars}
   \titlerunning{}
   \author{Glenn-Michael Oomen \inst{1,2} 
           \and 
           Hans Van Winckel \inst{1}
           \and
           Onno Pols \inst{2}
           \and
           Gijs Nelemans \inst{2,1}
         }
   \institute{Instituut voor Sterrenkunde (IvS), KU Leuven,
              Celestijnenlaan 200D, B-3001 Leuven, Belgium\\
              \email{glennmichael.oomen@kuleuven.be}
            \and
            Department of Astrophysics/IMAPP, Radboud University, P.O. Box 9010, 6500 GL Nijmegen, The Netherlands
   }
   \date{Received 8 May 2019 / Accepted 1 August 2019}
   \authorrunning{Oomen et al.}

  \abstract{
Many disc-type post-asymptotic giant branch (post-AGB) stars are chemically peculiar, showing underabundances of refractory elements in their photospheres that correlate with condensation temperature. The aim of this paper is to investigate how accretion from a circumbinary disc can cause this phenomenon of depletion and how this impacts the evolution of post-AGB stars. We used the \texttt{MESA} code to evolve stars in the post-AGB phase, while including accretion of metal-poor gas. We compared the models to a sample of 58 observed disc-type post-AGB stars with chemical abundance data. For each of these stars, we estimated the luminosity and the mass using the \textit{Gaia} distance. We modelled the accretion rate onto the binary from a viscously evolving disc for a range of initial accretion rates and disc masses. We find that large initial accretion rates ($\gtrsim 3\times10^{-7}$~$M_\sun$/yr) and large initial disc masses ($\sim10^{-2}$~$M_\sun$) are needed to reproduce the observed depleted post-AGB stars. Based on these high accretion rates, the evolution timescale of post-AGB stars can be significantly extended by a factor between two and five. We distinguish depletion patterns that are unsaturated (plateau profile) from those that are saturated, and we expect that post-red giant branch (post-RGB) stars are much more likely to show an unsaturated abundance pattern compared to post-AGB stars. Finally, because of the slower evolution of the low-mass post-RGB stars, we find that these systems can become depleted at lower effective temperatures ($<5000$~K). We conclude that accretion from a circumbinary disc successfully accounts for the chemical peculiarity of post-AGB stars.
}
  
 
   \maketitle
%

\section{Introduction} \label{sect:intro}
The post-asymptotic giant branch (post-AGB) phase is one of the last stages in the evolution of low- and intermediate-mass stars (0.8--8~$M_\sun$). Current evolutionary models of the post-AGB phase are still plagued by many uncertainties since any ambiguities in the previous evolutionary phases propagate to the final stages in stellar evolution. Moreover, the lack of empirical data on important physical properties, such as wind mass loss, hampers the models \citep{millerbertolami16}. Furthermore, the impact of binarity on post-AGB evolution is also still poorly understood \citep{demarco17}.

During the post-AGB phase, the star undergoes a rapid transition from the AGB to the planetary nebula (PN) phase. The post-AGB phase starts at the end of the dust-driven wind mass loss phase and ends when the star is hot enough to ionise its surroundings to become a planetary nebula at around 30\,000~K \citep{vanwinckel03}. During its evolution, the post-AGB star remains more or less at constant luminosity while increasing in effective temperature, and it therefore decreases in radius. During the post-AGB phase, the star shrinks from a radius of the order of an astronomical unit to the radius of the Sun at around 30\,000~K, and continues to shrink to about the radius of the Earth as it becomes a hot central star of a PN at 100\,000~K. 

The evolution of a post-AGB star is mainly determined by the amount of mass left in its envelope, as this is the main factor that determines the radius and hence temperature of the star \citep{vassiliadis94, blocker95}. The post-AGB phase can be divided into the following two parts: the transition timescale and the crossing timescale \citep{renzini89,vassiliadis94,marigo04,weiss09,millerbertolami16}. During the transition timescale, the post-AGB star loses the remainder of its already thin envelope via a stellar wind. This is the `slower' phase in post-AGB evolution since the effective temperature does not change much. At some point, the mass in the envelope of the post-AGB star becomes so low that a small change in envelope mass results in a large decrease in the radius of the star, and hence a large increase in effective temperature. This is called the crossing timescale since the star crosses the HR-diagram during this phase. 

Since post-AGB evolution is quite fast, material that is expelled at the end of the AGB phase can still be relatively close to the post-AGB star. This dusty gas is visible in the spectral energy distribution (SED) in the form of an infrared excess. If this excess peaks at mid-infrared wavelengths giving rise to a double-peaked SED, then the dust likely resides in a shell that is slowly receding away from the central star. If the excess starts at near-infrared wavelengths (at 2--3~$\mu$m), the dust is then still close to the star, which is an observational indication that the dust resides in a stable disc \citep[e.g.][and references therein]{hillen17, kluska18}. 

Disc-type post-AGB stars have been linked to binarity as all detected post-AGB binaries turned out to be disc sources \citep[e.g.][]{waelkens91,waelkens96,vanwinckel95,pollard95}. This is corroborated by the argument that circumbinary discs can form due to mass loss via the L$_2$ Lagrangian point \citep{shu79,frankowski07,pejcha16}. It is interesting to note that post-AGB stars with a disc-type SED are relatively common and, in our Galaxy alone, some 80 disc-type post-AGB stars are known \citep{deruyter06, gezer15, vanwinckel17}. Surveys by \citet{kamath14, kamath15} of post-AGB stars in the Magellanic Clouds have shown that the disc sources even outnumber post-AGB stars with a shell-type SED.

Since disc-type post-AGB stars are understood to be the result of a binary interaction, it is expected that a significant fraction of stars interact in a similar way on the red giant branch (RGB) rather than the AGB. Consequently, there are expected to be a large number of the so-called post-RGB stars \citep{kamath16} among the binary sample. These stars are observationally very similar to the post-AGB stars and can only be distinguished based on their luminosity, and in some cases, surface abundances. This is due to the strong relation between the core mass and the luminosity of these objects \citep{vassiliadis94}. In contrast to the CO cores of post-AGB stars, post-RGB stars have degenerate helium cores since the core mass is too low to ignite helium. These stars eventually evolve to become helium white dwarfs. Throughout this work, we simply refer to the combined group of post-RGB and post-AGB stars as the post-AGB stars, unless specified otherwise.

An interesting observational property of post-AGB stars is that a large fraction of these stars are chemically peculiar, showing in their photospheres an underabundance in refractory elements, a phenomenon called depletion \citep{vanwinckel95,vanwinckel97}. The underabundance often scales with the condensation temperature of a particular element \citep{maas05,giridhar05}. The phenomenon of depletion is closely related to the presence of a disc, since all depleted post-AGB stars are disc sources \citep{gezer15}. The suggested mechanism for depletion is re-accretion of gas from a circumbinary disc, while the dust, which contains most of the refractory elements, feels a much larger radiation pressure and hence does not get accreted \citep{waters92}. A disc is a necessary condition for depletion, but not sufficient, as not all post-AGB stars with discs are depleted \citep[e.g.][]{gezer15}. Recent work by \citet{oomen18} on a sample of post-AGB binaries shows that only post-AGB stars with high temperatures ($T_{\mathrm{eff}} > 5000$~K) in relatively wide orbits ($a_1 \gtrsim 0.3$~AU) are depleted.

Evidence for accretion from a circumbinary disc to the inner binary is becoming more abundant, since many post-AGB binaries show blue-shifted absorption by a jet at superior conjunction, when an accreting companion moves in front of the post-AGB star \citep{gorlova15,hillen16,bollen17}. Since Roche-lobe overflow is unlikely due to the decreasing radius of the post-AGB star and a strong stellar wind is not observed in these systems, the jet is likely created by an accretion disc around the companion, which is fed by the circumbinary disc. Moreover, some post-AGB binaries show a double-peaked H$\alpha$ profile following the radial velocity motion of the post-AGB star, which could be indicative of an accretion disc around the post-AGB star \citep{gorlova12}.

Because post-AGB evolution is strongly dependent on the amount of mass in the hydrogen-rich envelope of the star, accretion can extend the post-AGB evolution as new fuel is added to the star \citep{izzard18}. However, this topic has not been investigated quantitatively and few details are known. The purpose of this contribution is to investigate the impact of accretion on the evolution of post-AGB stars. We evaluate whether gas accretion from a circumbinary disc is able to produce the observed photospheric depletion and we constrain the model parameters by comparing our models to a sample of observed post-AGB stars.

Understanding the impact of discs and binarity on post-AGB evolution is of major importance for the formation and morphology of PNe, which depend strongly on the post-AGB timescale and mass-loss history \citep{millerbertolami16}. If the post-AGB star evolves too slowly, most of the material expelled at the end of the AGB phase will have dispersed before the central star reaches ionising temperatures \citep{gesicki18}. Moreover, having some constraints on the accretion properties of the gas will improve our understanding of the complex circumbinary morphologies of the post-AGB binary systems \citep[e.g.][]{hillen14, hillen15,hillen17,kluska18}. Finally, a thorough understanding of the evolution of binary post-AGB stars can provide important clues to the poorly understood binary interactions in the AGB phase \citep{demarco17, jones17b}.

In Sect.~\ref{sect:data}, we present the observational data that is used in this work. In Sect.~\ref{sect:gasdepletion}, we discuss a simple model of gas dilution and compare the general shape of expected depletion patterns to those observed in nature. In Sect.~\ref{sect:methods}, we derive luminosities for stars in our sample and we give details on the \texttt{MESA} models used in this work. We present the results in Sect.~\ref{sect:results}. Finally, we discuss our findings in Sect.~\ref{sect:discussion}.

\section{Data} \label{sect:data}
We collected spectroscopic data for all known disc-type post-AGB stars in our Galaxy with chemical abundances available in literature. This resulted in a total of 58 post-AGB stars in our sample. These stars are presented in Table~\ref{CHEMDATA} of appendix~\ref{appendix:data} along with the effective temperatures, tracers for depletion ([Zn/Ti], [Zn/S], and [Zn/Fe]), and some general characteristics of the depletion pattern which will be discussed in Sect.~\ref{sect:depletionpatterns}. In addition to spectroscopic data, we collected distances from the catalogue of \citet{bailerjones18} for all our targets (see second column of Table~\ref{LUMDATA}). 

\section{Gas depletion} \label{sect:gasdepletion}

\subsection{Observed depletion patterns} \label{sect:depletionpatterns}

\begin{figure*}
\resizebox{\hsize}{!}{\includegraphics{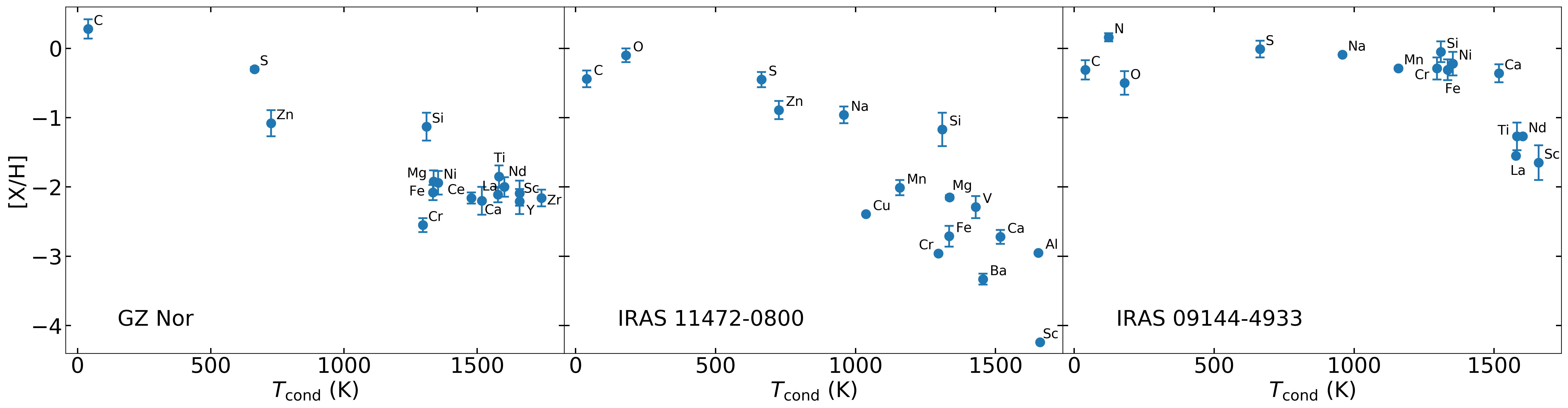}}
\caption{Depletion patterns of GZ~Nor \citep{gezer19} (left), IRAS~11472-0800 \citep{vanwinckel12} (middle), and IRAS~09144-4933 \citep{maas05} (right). The abundances displayed are with respect to solar composition. There is some diversity in the shape of the patterns, but the general trend is a decrease of the element abundance with respect to condensation temperature. Condensation temperatures are taken from \citet{lodders03} and are defined as the equilibrium temperature at which half of an element has condensed into dust at constant gas pressure.}
\label{fig:observedpatterns}
\end{figure*}

Depletion is the systematic underabundance of refractory elements in the photospheres of stars. The higher the condensation temperature of a particular element, the lower its observed abundance \citep{maas05, giridhar05, vanwinckel12}. However, observations show not just one relation between condensation temperature and abundance, but a plethora of different depletion patterns. Several examples of this diversity are shown in Fig.~\ref{fig:observedpatterns}. 

Generally, three different factors describe the shape of a depletion pattern. First, there is the depletion value, which is the ratio of the abundance of one of the most depleted elements (Ti or Sc) to a non-depleted, volatile element (Zn or S). This ratio is always written as compared to solar composition. The [Zn/Ti] ratio is most commonly used as a tracer for how depleted a star is, because these two elements are at the iron peak which means they share a similar nucleosynthesis history, keeping intrinsic abundance scatter to a minimum. Moreover, Ti has good, unblended lines to use for chemical analysis \citep[see e.g.][]{jofre18,laverick19}.

The second factor is the turn-off temperature, which defines the limit for which elements with higher condensation temperatures become depleted. For GZ~Nor and IRAS~11472-0800 (Fig.~\ref{fig:observedpatterns}, left and middle panel), this temperature is around 800--1000~K, since elements like Na and Mn (and even Zn) have underabundances compared to solar. This is in contrast to IRAS~09144-4431 (Fig.~\ref{fig:observedpatterns}, right panel), where only the most refractory elements with condensation temperatures larger than 1500~K are depleted.

The third factor describes the overall shape of the profile. In most cases, the underabundance scales with condensation temperature, because elements such as Ti are more depleted than Fe or Mg. However, for a small fraction of the depleted post-AGB stars, the underabundance is approximately the same for all refractory elements, despite the condensation temperatures of those elements being different. A clear case of such a profile is GZ~Nor, where the abundance of Ti and Sc are the same as those of Fe and Mg.

\subsection{Gas dilution} \label{sect:gasdilution}

As gas is accreted onto a star, the gas mixes with the convective outer layers. If the chemical composition of the accreted gas differs from that of the star, the chemical composition in the photosphere will gradually change as more gas is accreted, which results in the observed depletion patterns of post-AGB stars. 

In order to determine the chemical composition of the accreted material, we used the observed chemical abundances in depleted post-AGB objects as a basis. If we assume that the chemical composition of the accreted material is the same for all post-AGB stars, then we can estimate this composition based on both the condensation temperature of an element as well as its observed abundance in some of the most depleted post-AGB stars.

\begin{figure}
\resizebox{\hsize}{!}{\includegraphics{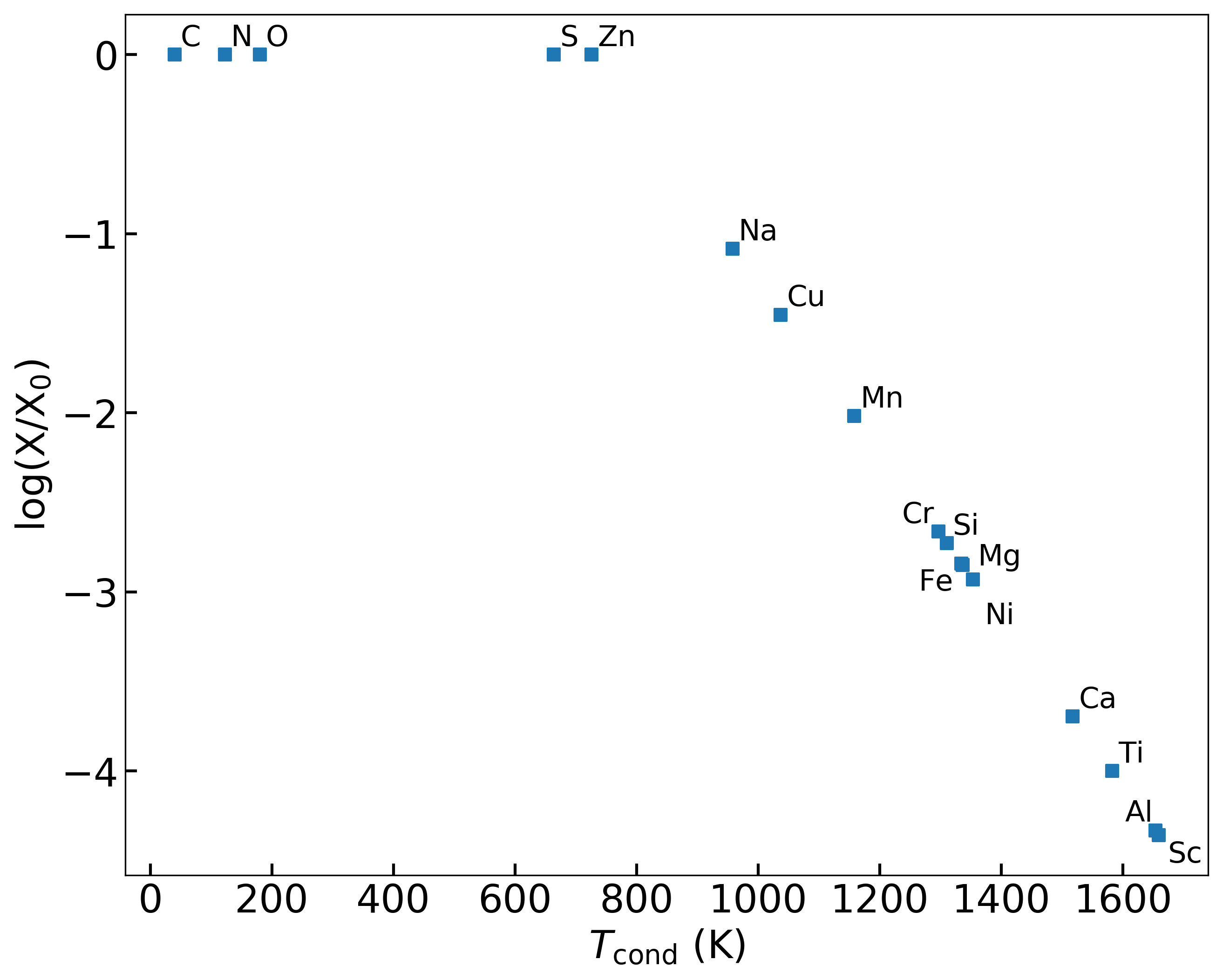}}
\caption{Estimate for chemical abundances of accreted gas. This is based on condensation temperatures of elements and the most depleted observed post-AGB stars. The abundances displayed are with respect to the initial composition of the star. Elements with higher condensation temperatures ($T_\mathrm{cond}$) have lower accretion abundances. We use this chemical composition for the accreted gas in our \texttt{MESA} models.}
\label{fig:chemabun}
\end{figure}

IRAS~11472-0800 (middle panel of Fig.~\ref{fig:observedpatterns}) is a good example of one of the most depleted post-AGB stars \citep{vanwinckel12}. Since the abundances of the most refractory elements, such as Sc, are around $-4$~dex relative to solar, we take the accretion abundance of Ti to be $-4$~dex. Furthermore, we keep the abundances of the volatile elements until Zn to be 0~dex, since this element shows no underabundance in most post-AGB stars. The underabundance of elements with respect to condensation temperature has an approximately linear relation, albeit with a lot of scatter. This scatter is likely  not only due to errors in determining the abundances, but also because the condensation temperatures are determined for constant gas pressure in equilibrium \citep{lodders03}. It is very likely that there are star-to-star intrinsic differences in the abundances of the accreted gas. However, taking a linear relation for the underabundance of an element with respect to its condensation temperature, as well as fixing the abundance of Zn and Ti to 0~dex and $-4$~dex, respectively, yields the abundance pattern shown in Fig.~\ref{fig:chemabun}.

\begin{figure*}
\resizebox{\hsize}{!}{\includegraphics{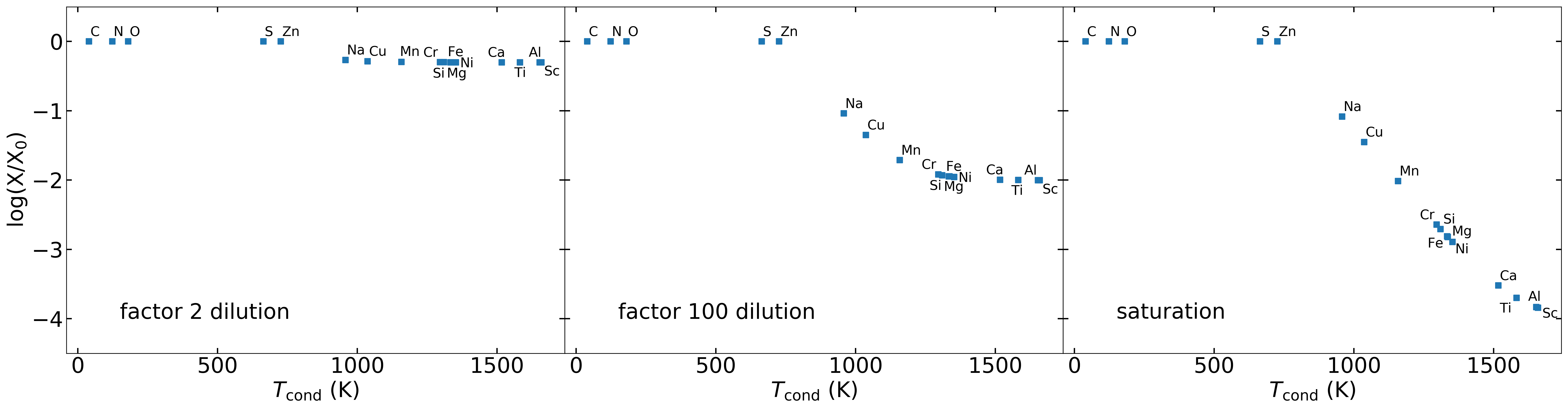}}
\caption{Chemical abundances of a gas mixture after dilution of a solar-composition gas with abundances from Fig.~\ref{fig:chemabun}. The left panel shows the composition after dilution by a factor of two. In the middle panel, the gas is diluted by a factor of 100. Finally, in the right panel, the gas is diluted by a factor of 10\,000, to the point where we retrieve the initial accretion abundances from Fig.~\ref{fig:chemabun}.}
\label{fig:gasdil}
\end{figure*}

We illustrate the effect of gas dilution in Fig.~\ref{fig:gasdil}. We dilute a gas with initial abundances equal to 0~dex with metal-poor gas using abundances from Fig.~\ref{fig:chemabun}. The left panel shows the abundances of the gas when the amount that has been accreted is equal to the initial mass of the mixture. Since the refractory elements in the accretion composition are depleted, the abundances of the those elements will decrease by a factor of two (0.3~dex). In the middle panel of Fig.~\ref{fig:gasdil}, we show the abundance pattern when the gas has been diluted 100~times. Now, the vast majority of the mixture consists of accreted gas, which means that metals that have mild underabundances in the accreted gas (Na and Cu) become saturated and converge to the abundance of the accreted gas in Fig.~\ref{fig:chemabun}. Metals that have accretion abundances [X/H] less than $-2$~dex will converge to $-2$~dex as still 1$\%$ of the mixture consists of the initial gas. Finally, the right panel shows the saturated mixture after diluting more than 10\,000~times, such that even the elements with the highest condensation temperatures converge to their accretion abundances. At this point, we retrieve the chemical composition of the accreted gas from Fig.~\ref{fig:chemabun}.

By comparing the chemical composition of different stages of diluted gas from Fig.~\ref{fig:gasdil} with the observed depletion patterns of post-AGB stars in Sect.~\ref{sect:depletionpatterns}, we can easily identify that the `saturated' profile in the right panel of Fig.~\ref{fig:gasdil} is quite similar to the pattern of IRAS~11472-0800 in Fig.~\ref{fig:observedpatterns}. Furthermore, the middle panel of Fig.~\ref{fig:gasdil} shows a `plateau' of refractory elements with similar underabundances, much like that of GZ~Nor in the left panel of Fig.~\ref{fig:observedpatterns}. 

By using this simple gas-dilution model, we better understand the difference between the plateau and saturated profiles of depleted post-AGB stars. However, the observed diversity regarding the maximum depletion value and the turn-off temperature cannot be reproduced, as is the case for example for IRAS~09144-4933 in the right panel of Fig.~\ref{fig:observedpatterns}. This indicates that the chemical composition of the accreted gas can differ from star to star. For example, since the Ti abundance for IRAS~09144-4933 is $-1.3$~dex, we know that the atmosphere of this star must have been diluted at least by a factor of 20, yet the observed abundance of Fe is only $-0.3$~dex. This means that the accretion abundance of Fe should also be $-0.3$~dex, which is not the case for some of the more depleted stars in the sample. 

\begin{figure}
\resizebox{\hsize}{!}{\includegraphics{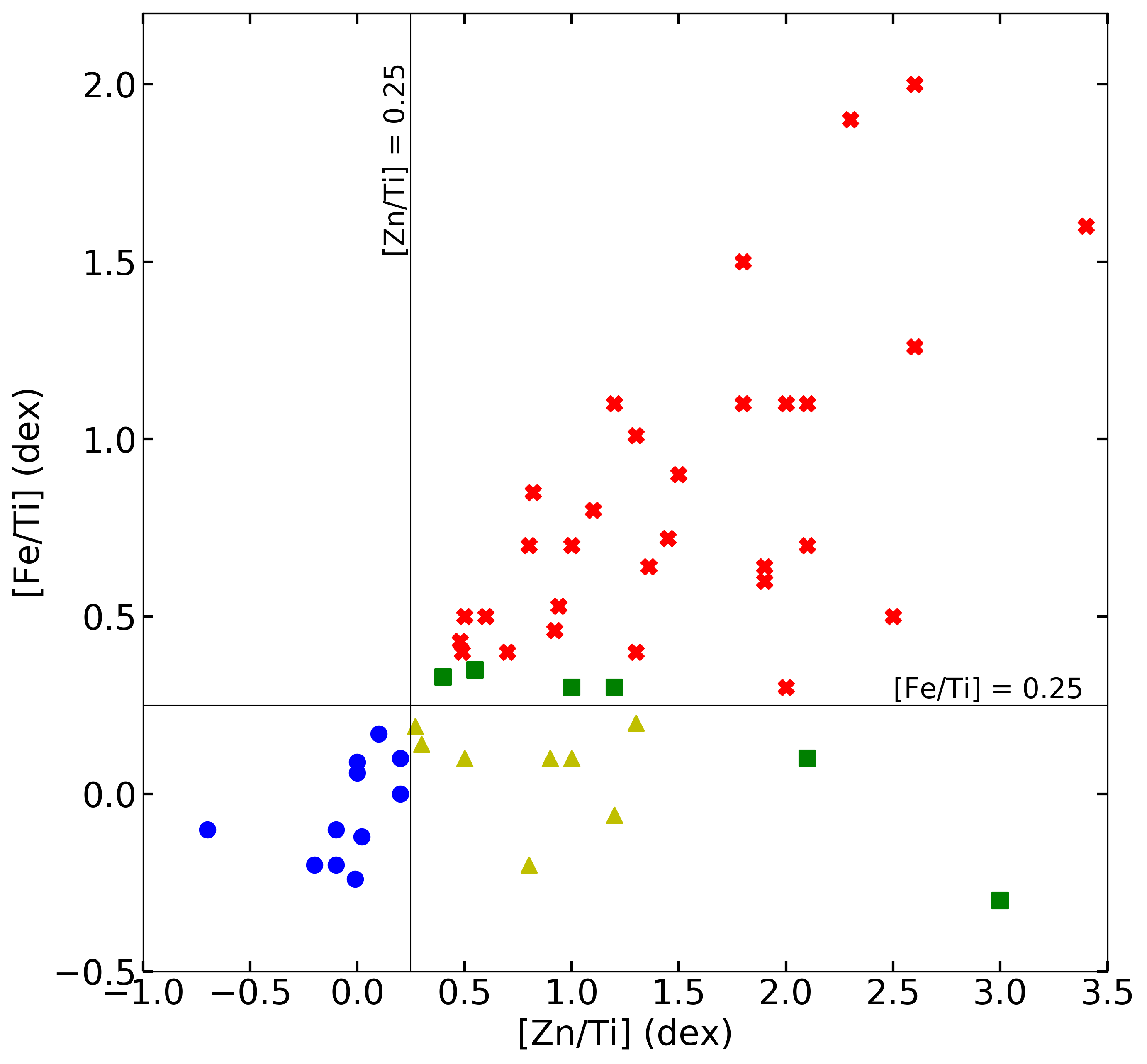}}
\caption{Observed abundances of stars in our sample plotted in the [Zn/Ti]--[Fe/Ti] plane. The [Zn/Ti] tracer defines whether stars are depleted, while [Fe/Ti] distinguishes between saturated profiles and plateau profiles. Non-depleted objects are plotted as blue circles, plateau profiles are shown as yellow triangles, and saturated objects are shown as red crosses. Objects with uncertain depletion patterns are denoted by green squares. Objects without a Ti (or Sc) measurement (CC~Lyr, HD~44179, and HR~4049) have been excluded from this figure.}
\label{fig:patterndiagram}
\end{figure}

We present the (approximate) turn-off temperatures for all stars in Table~\ref{CHEMDATA} and assign whether these stars have a plateau profile (P) or a saturated profile (S). Figure~\ref{fig:patterndiagram} shows the distribution of the different patterns in a [Fe/Ti]--[Zn/Ti] diagram. The [Zn/Ti] tracer shows how depleted a star is, while the [Fe/Ti] ratio traces whether a star is saturated ([Fe/Ti] > 0.25) or not saturated ([Fe/Ti] < 0.25). The green squares denote stars with uncertain patterns (U). This can be either because the star has a pattern in between plateau and saturation, or because the star is so depleted that even Fe has an extremely low abundance. When the element abundance becomes less than $-4$~dex, it becomes hard to detect any spectroscopic lines. We find the non-depleted (N) objects in the lower left corner of Fig.~\ref{fig:patterndiagram}. There is one outlier in our sample that has [Zn/Ti] $=-0.7$~dex, which cannot be reproduced by our dilution model. This object, DF~Cyg, either has a peculiar Zn or Ti abundance, or this might be due to an observational error in the abundance determination.

A noticeable feature in Fig.~\ref{fig:patterndiagram} is that most post-AGB stars exhibit saturated-like depletion patterns, but for a whole range of [Zn/Ti] ratios, suggesting that the composition of the accreted gas is less depleted than in Fig.~\ref{fig:chemabun}. Additionally, only a few post-AGB stars show plateau-like features. This suggests that the photospheres of those stars have not been diluted to the extent that the chemical abundances that are observed are the same as those of the accreted gas. This will be discussed in more detail in Sect.~\ref{sect:plateauvssaturation}.

The large range in turn-off temperatures observed in post-AGB stars also shows that there is a large diversity in composition of accreted gas, since gas dilution does not change the turn-off temperature of a gas mixture. Consequently, defining only one composition of the accreted gas for the whole population of post-AGB and post-RGB stars does not allow for quantitative modelling of the full abundance patterns of individual post-AGB stars. Instead, by using the [Zn/Ti] ratio, we can still trace the level of dilution in the post-AGB photosphere. This will be discussed in Sect.~\ref{sect:comparingmodels}.

\section{Methods} \label{sect:methods}

\subsection{Mass estimation} \label{sect:massestimation}
We aim to describe the formation of depletion patterns in post-AGB stars in more detail by computing a set of detailed stellar-evolution models while including accretion of depleted gas. These models are then compared to observed post-AGB stars of similar mass. Consequently, we first estimated the masses of observed post-AGB stars based on their luminosity.

\subsubsection{Luminosity determination} \label{sect:luminositydetermination}
To determine the luminosities of the stars in our sample, we have fitted the SEDs and subsequently integrated the photospheric contribution. The SED fitting procedure has been described in previous works \citep[see e.g.][]{manick17,oomen18}. Since most of our sources show a near-infrared excess due to the presence of hot dust in a circumbinary disc, we fitted only the photometry of the SED at wavelengths where the photospheric contribution of the post-AGB star dominates while correcting for both interstellar and circumstellar reddening. Here we assume that the total extinction follows the wavelength dependence of the interstellar-medium extinction law \citep{cardelli89}.

Once we derive the integrated flux, we can easily compute the intrinsic luminosity of the star using the distances from \citet{bailerjones18}. These distances are based on the second data release of \textit{Gaia} \citep{gaiadr2}, hence are based on single-star astrometric solutions even though we expect all stars in our sample to be binary systems \citep{gaiadr2astrometric,gaiadr2parallax}. This results in unreliable distances for some of our stars. Nevertheless, the distances from \textit{Gaia} have been shown to agree fairly well with other distance measurements in some cases, such as IRAS~08544-4431 \citep{kluska18}, RV~Tau and DF~Cyg \citep{manick19}, and GK~Car and GZ~Nor \citep{gezer19}. For other cases, the parallax measurement from \textit{Gaia} is highly likely to be affected by the binary motion. For example, the \textit{Gaia} parallax for HR~4049 is $0.58\pm0.15$~mas, while the Hipparcos parallax gives $1.50\pm0.63$~mas, which would (within 1$\sigma$ error) result in a much smaller distance and a much lower luminosity for this star. Moreover, for some RV~Tauri pulsators such as DY~Ori and EP~Lyr, the luminosity based on parallax measurements disagrees with that of the period-luminosity relation by a factor of five to ten (2--3$\sigma$) \citep{manick17}. Nonetheless, we only use distances determined from the \textit{Gaia} parallax in this work for sake of consistency. All distances and luminosities, as well as the ratio of luminosity from the infrared excess to the photospheric luminosity, are given in Table~\ref{LUMDATA}.

\subsubsection{Mass assignment} \label{sect:massassignment}

The main property of a (post-)AGB star that determines its luminosity is the core mass \citep{vassiliadis94,blocker95}. The more massive the CO- or He-core, the more luminous the star. Consequently, we can use the luminosity of a star to estimate its current mass. By separating the stars into several mass bins, we can compare our \texttt{MESA} models to the observed sample in our results. 

Based on the luminosities of the \texttt{MESA} models (Sect.~\ref{sect:mesamodels}) for different core masses, we defined bins into which we sorted the observed post-RGB and post-AGB luminosities. These are chosen to correspond to three groups: likely post-RGB stars, likely low-mass post-AGB stars, and likely high-mass post-AGB stars. If the luminosity of a given object is less than 2500~$L_\sun$, we classify it as a 0.40~$M_\sun$ post-RGB star. A star is classified in the 0.55~$M_\sun$ bin if it has a luminosity in the range 2500--7500~$L_\sun$. Finally, stars with luminosities larger than 7500~$L_\sun$ are classified as a 0.65~$M_\sun$ star. 

In our binning procedure, we rejected all objects for which the ratio of the infrared-to-photospheric luminosity is larger than two. This is because if the luminosity of the disc is much larger than the luminosity of the star, we are observing the system close to edge-on. In those cases, the light of the star is partly obscured, which means that we greatly underestimate the intrinsic luminosity of the star. Furthermore, we note that, as the \textit{Gaia} parallaxes are based on single-star models, some outliers with very large and very small values are to be expected; this results in an apparent overabundance of objects in the 0.40 and 0.65~$M_\sun$ bins, respectively.

\subsection{MESA models} \label{sect:mesamodels}
We used version 10398 of the stellar evolution code \texttt{MESA} \citep{MESApaper1,MESApaper2,MESApaper3,MESApaper4} to model the effect of accretion on the post-AGB phase. We used the single-star module to evolve a star through the post-AGB phase, while including accretion of gas poor in refractory elements onto the post-AGB star. Since the evolution of a post-AGB star is strongly dependent on the core mass, we modelled the evolution for a range of core masses: 0.40, 0.45, 0.55, 0.60, and 0.65~$M_{\sun}$.\footnote{When the star has a core mass of 0.50~$M_\sun$, it is in a core helium burning stage, which means it cannot become a post-RGB/AGB star.} 

The post-RGB models ($M < 0.47 M_\sun$) are prepared by evolving a star with initial mass of 1.5~$M_\sun$ and solar metallicity \citep{asplund09} to the RGB phase. Once the helium core reaches a mass of either 0.40~$M_\sun$ or 0.45~$M_\sun$, we artificially increase the mass-loss rate to $10^{-4}$~$M_\sun$/yr, effectively removing the hydrogen-rich envelope on a timescale of $\sim$10\,000~yrs. For the post-AGB models, we apply the same procedure, but instead use an initial star mass of 2.5~$M_\sun$ and evolve it to the AGB phase. Furthermore, we start removing the envelope at the point where the CO core + He shell mass reaches 0.55~$M_\sun$, 0.60~$M_\sun$, and 0.65~$M_\sun$. 

We remove the envelope to the point at which only 0.015~$M_\sun$ and 0.02~$M_\sun$ of the hydrogen-rich envelope was left for the post-RGB and post-AGB models, respectively. Here, the star has started its evolution towards higher temperatures. At that point, we do not differentiate between post-RGB and post-AGB and we simply call this post-AGB evolution. We start our post-AGB evolution models at this point, varying a set of parameters in the accretion model discussed below, including the effective temperature at which accretion starts (Sect.~\ref{sect:parameters}).

\subsubsection{Input physics} \label{sect:inputphysics}
Within \texttt{MESA}, we used a preset nuclear reaction network designed to evolve AGB stars and so-called super-AGB stars, the more massive analogues of AGB stars. This network contains 51 nuclear reactions with 27 isotopes of elements up to \element[][]{Al} participating. In addition, we included several other elements to the network in order to model the depletion patterns observed in post-AGB stars. These elements include \element[][28]{Si}, \element[][32]{S}, \element[][40]{Ca}, \element[][45]{Sc}, \element[][48]{Ti}, \element[][52]{Cr}, \element[][55]{Mn}, \element[][56]{Fe}, \element[][58]{Ni}, \element[][63]{Cu}, and \element[][64]{Zn}.

Convection is treated with the mixing length theory, which uses a free parameter called the mixing length, $\alpha_{\mathrm{ML}}$. This parameter can be calibrated empirically with asteroseismology, although its value is still difficult to constrain and likely depends on some stellar properties, such as metallicity and evolutionary state \citep{tayar17, li18}. For that reason, we simply take $\alpha_{\mathrm{ML}} = 2$ throughout this work. Even though we expect convective mixing to dominate, we also allow for semiconvection and thermohaline mixing \citep{MESApaper2}. Furthermore, we assume that the stars are non-rotating, since these objects are observed to be slowly rotating.

It is important to note that we do not include convective overshooting in our models. Although it plays an important role in stellar evolution, especially for the initial--final mass relation, we found that including convective overshooting hampers the stability of the code during the post-AGB phase in the more massive models. Since convective overshooting only mildly affects the stellar parameters during the post-AGB phase and does not significantly increase the extent of mixing in the outer layers, we decided to exclude this from all models altogether.

Stellar winds are an important element in post-AGB evolution, since they directly impact on the envelope mass of the star. Unfortunately, empirical data on stellar winds during the post-AGB phase are lacking. However, lower-temperature post-AGB stars are very similar to RGB stars in the sense that they are high-luminosity objects with an extended, deep convective envelope. For that reason, we used a semi-empirical RGB-type wind prescription from \citet{schroder05,schroder07} who have updated the Reimers law \citep{reimers75} to include a dependence on effective temperature and surface gravity:
\begin{equation}
\dot{M} = 8\times10^{-14} M_{\sun}/\mathrm{yr} \frac{L_* R_*}{M_*}\left(\frac{T_{\mathrm{eff}}}{4000 \mathrm{K}}\right)^{3.5}\left(1+\frac{g_\sun}{4300g_*}\right).
\label{eq:wind}
\end{equation}
We note that the prescription of \citet{schroder05} is based on stellar winds driven by chromospheric activity, which is not observed in post-AGB stars. Furthermore, the wind mass-loss mechanism depends strongly on the magnetic field strength, which is observed to be weak or non-existent in our objects \citep{sabin15}. \citet{cranmer11} show that this mechanism is unable to properly model post-AGB winds. Nonetheless, since there is no better alternative, we used the prescription of \citet{schroder05} for the lower-temperature stages of post-AGB evolution.

To describe stellar winds at higher temperatures, we used the prescription given by \citet{millerbertolami16}:
\begin{equation}
\dot{M} = 9.778\times10^{-15} M_{\sun}/\mathrm{yr} \left(\frac{L_*}{L_\sun}\right)^{1.674} \left(\frac{Z_0}{Z_\sun}\right)^{2/3}.   
\end{equation}
This formula describes a radiation-driven wind calibrated for central stars of planetary nebulae (CSPNe). Even though this formula only applies to objects hotter than 30\,000~K, it yields mass-loss rates of the order of $10^{-9}$~$M_\sun$/yr, which is much lower than the rate at which hydrogen is burned into helium at the bottom of the convective envelope ($\sim 10^{-7} M_\sun$/yr). Consequently, the impact of stellar winds in hotter post-AGB stars and CSPNe on their evolution is negligible.

Similar to \citet{millerbertolami16}, we defined a transition region in which the wind transforms from the cool RGB wind regime to the hot CSPN wind regime. This transition is arbitrary due to the lack of observational constraints. In this work, we have chosen to exponentially transition from the RGB wind to the CSPN wind between $\log T_\mathrm{eff} = 3.7$ and $\log T_\mathrm{eff} = 3.9$, since this is the temperature at which the wind as described by Eq.~\ref{eq:wind} is expected to disappear \citep{linsky79, ayres10, cranmer11}.

\subsubsection{Accretion} \label{sect:accretion}
Our goal is to investigate how accretion of gas from a circumbinary disc impacts post-AGB evolution. In early work, it was expected that the effect of the binary torque on a circumbinary disc would impair accretion onto the central binary, as the central cavity is rapidly cleared of material up to a radius of about twice the binary separation \citep{pringle91, artymowicz94}. However, more recent 2D and 3D simulations show accretion streams that penetrate the inner gap leading to accretion rates that are as high as or even higher than rates of models without a binary torque \citep{macfadyen08,shi12,dorazio13,farris14,shi15}. In order to simulate the accretion rate onto the central binary, we followed \citet{rafikov16} for the viscous evolution of a circumbinary disc.

In this prescription, the accretion rate into the binary cavity scales with the density in the disc. As the disc evolves, the density in the disc slowly decreases as the outer rim of the disc expands and the central binary accretes gas, effectively decreasing the total disc mass. \citet{rafikov16} gives the following expression for accretion without a central binary torque:
\begin{equation}
\dot{M}(t) = \frac{M_\mathrm{d}(0)}{2t_0}\left(1+\frac{t}{t_0}\right)^{-3/2},
\label{eq:accrrafikov}
\end{equation}
where $M_\mathrm{d}(0)$ is the initial disc mass and $t_0$ is the initial viscous time of the disc given by
\begin{equation}
t_0 = \frac{4}{3}\frac{\mu}{k_B}\frac{a_\mathrm{b}}{\alpha}\left(\frac{4\pi\sigma(GM_\mathrm{b})^2}{\zeta L_*}\right)^{1/4}\left(\frac{\eta}{I_L}\right)^2,
\label{eq:t0rafikov}
\end{equation}
where $\mu$ is the mean atomic weight, $a_\mathrm{b}$ is the binary separation, $\alpha$ is the viscosity parameter, $M_\mathrm{b}$ is the binary mass, and $L_*$ is the luminosity of the post-AGB star. $\zeta$ is a geometrical factor for the incidence angle of stellar radiation on the disc. Finally, $\eta$ is the ratio of the specific angular momentum of the disc to that of the binary, while $I_L$ is a constant that describes the distribution of angular momentum in the disc \citep{rafikov16b}.

By taking typical values for a post-AGB binary with $L_* = 5000$~$L_\sun$, $a_\mathrm{b} = 2$~AU, and $M_\mathrm{b} = 1.7$~$M_\sun$ \citep{oomen18}, and assuming a disc mass of $10^{-2}$~$M_\sun$ \citep{hillen17, kluska18}, we can derive the typical accretion rate onto a post-AGB binary. Here, we take $\zeta = 0.1$, $\eta = 2$, $\alpha = 0.01$, and $I_L = 1$ \citep[see][]{rafikov16}. We find a typical accretion rate of $\sim 10^{-7}$~$M_\sun$/yr, which is in agreement with the accretion rate in the model of \citet{izzard18}.  

We choose the same chemical composition for the accreted gas as that given in Fig.~\ref{fig:chemabun} of Sect.~\ref{sect:gasdilution}. Here, we take the abundances in the outermost layers of the star at the start of the post-AGB phase as the initial composition. This implies that if the abundances have changed with respect to the initial solar composition due to the AGB nucleosynthetic history, the accreted abundances also change with respect to solar. Consequently, the abundances of elements such as carbon do not change as a result of accretion. We note that some refractory $s$-process elements can be significantly enhanced during the AGB phase, but we do not take these into account in the model.

The depleted gas is added over time to the outermost layers of the star. To avoid accumulating all the gas in the outer cells of the \texttt{MESA} model and thereby instantly creating an optically thick layer with the accreted gas composition, we imposed some mixing in the outer layers of the star. In a convective envelope, turbulent motions cause the accreted material to be mixed very rapidly. However, as the post-AGB star becomes hotter, the large convective envelope starts to fragment into smaller convective layers with large opacity. To keep the accreted material mixed with the outer envelope for the higher temperature post-AGB stars ($>5000$~K), we imposed that the outer layer is mixed down to at least the \ion{He}{ii} convective region at around 50\,000~K (see discussion in Sect.~\ref{sect:atmosphericmixing}).

\subsubsection{Parameters} \label{sect:parameters}

We model the time-dependent accretion rate in the post-RGB/AGB phase by using two separate parameters: the initial accretion rate ($\dot{M}(t=0)$) and the initial disc mass ($M_\mathrm{d}$). Furthermore, we assume that the post-AGB star only accretes approximately half of the gas entering the binary cavity, while the other half flows to the companion \citep[see Fig.~7 of][for the individual accretion rates in binaries with different mass ratios]{farris14}. This means that the disc mass loss rate is twice as large as the post-AGB accretion rate, such that $t_0 = M_\mathrm{d}/4\dot{M}(0)$ in Eq.~\ref{eq:accrrafikov}. The mass accretion rate onto the post-AGB star then becomes
\begin{equation}
\dot{M}_\mathrm{accr}(t) = \dot{M}(0)\left(1+\frac{4\dot{M}(0)t}{M_\mathrm{d}}\right)^{-3/2}.
\label{eq:accr}
\end{equation}
Since typical disc mass loss rates in post-AGB binaries are around $10^{-7}$~$M_\sun$/yr, we take five different accretion rates for $\dot{M}(0)$ in the range $5\times10^{-9}$--$5\times10^{-7}$~$M_\sun$/yr in steps of 0.5~dex in log-space, such that the rate of gas flowing into the binary cavity corresponds to $10^{-8}$--$10^{-6}$~$M_\sun$/yr. For the total initial disc mass $M_\mathrm{d}$, we use $10^{-3}$, $3\times10^{-3}$, and $10^{-2}$~$M_\sun$, which corresponds to the range of observed post-AGB disc masses \citep[e.g.][]{bujarrabal13}.

\begin{figure}
\resizebox{\hsize}{!}{\includegraphics{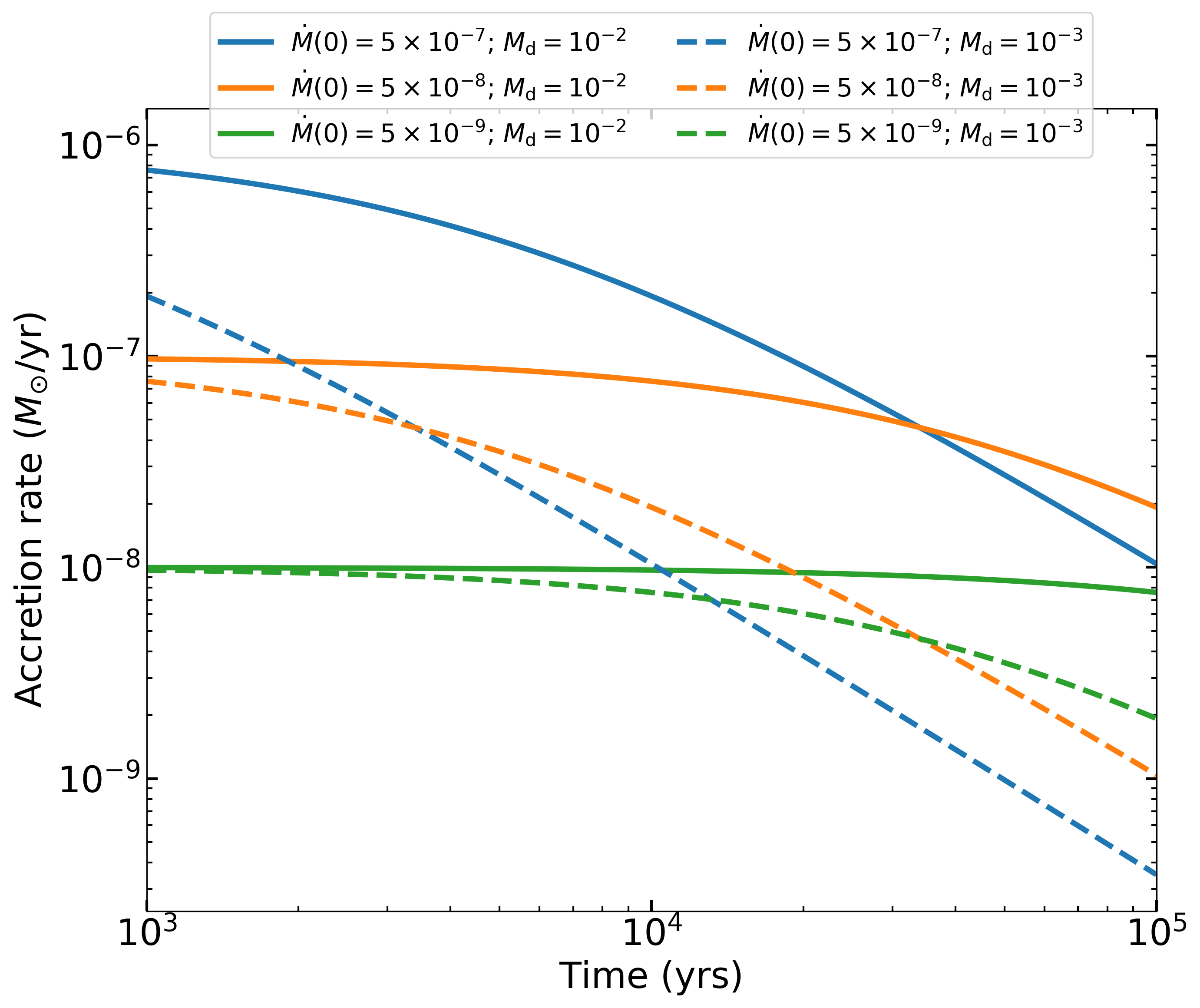}}
\caption{Evolution of accretion rates over time from the accretion model of Eq.~\ref{eq:accr}. Colours denote different initial accretion rates $\dot{M}(0)$ in units of $M_\sun$/yr, while different linestyles represent different initial disc masses $M_\mathrm{d}$ in units of $M_\sun$. Models with lower disc masses decrease faster over time.}
\label{fig:accretionrates}
\end{figure}

Figure~\ref{fig:accretionrates} shows the evolution of the accretion rate over time for several parameter choices in Eq.~\ref{eq:accr}. The figure reveals that low disc-mass models (dashed lines) decrease in accretion rate much faster compared to high disc-mass models (solid lines). This is a consequence of the evolution of the density in the disc, which reduces much faster for low disc-mass models paired with high accretion rates. In fact, the accretion rate of the high $\dot{M}(0)$--low $M_\mathrm{d}$ model (blue dashed line in Fig.~\ref{fig:accretionrates}) reaches the lowest accretion rate across all models after little over 10\,000~yrs. 

Another important parameter in modelling the accretion onto a post-AGB star is the moment at which the accretion starts. The circumbinary disc is presumably created from mass loss by an AGB star via the L$_2$ Lagrangian point of the binary system. Suppose that the phase of strong interaction stops as soon as the radius of the AGB star decreases due to the low envelope mass, then the star will start its post-AGB evolution while being close to filling its Roche lobe. Consequently, depending on the binary separation and hence the Roche-lobe radius, the post-AGB star can start its evolution at different radii. We mimic this effect by starting the accretion in the model at different effective temperatures: 3500~K, 4000~K, 5000~K, and 6000~K. Because of the tight relation between $T_\mathrm{eff}$ and envelope mass, this is equivalent to starting accretion at different envelope masses \citep[see Fig.~C.1 in][]{oomen18}. Furthermore, since we allow the star to evolve slowly towards a higher $T_\mathrm{eff}$ before we initiate accretion, we implicitly assume that the star leaves the binary interaction in thermal equilibrium. This is a reasonable assumption, because the Kelvin-Helmholtz timescale is only of the order of decades, which means that such a star will regain thermal equilibrium almost instantly after the interaction with the companion has stopped.


More information about the models will be made publicly available at the \texttt{MESA} marketplace at \verb|cococubed.asu.edu/mesa_market/inlists.html|. This includes \texttt{MESA} inlists and other source files used in this article, as well as a few modifications made to the source code to allow for simultaneous mass gain by accretion and mass loss by stellar winds, where the accreted gas has our specified chemical composition.

\subsection{Comparing models to observations} \label{sect:comparingmodels}

We assess how fast stars can become depleted by comparing effective temperatures to a tracer for depletion. We use the [Zn/Ti] ratio to characterise the depletion value for each star. In case we have no abundance data for Zn, we use the [S/Ti] ratio instead, while if there is no Ti (or Sc) abundance available for the star, we take [Zn/Fe] as the depletion value (see Table~\ref{CHEMDATA}). 

\begin{figure}
\resizebox{\hsize}{!}{\includegraphics{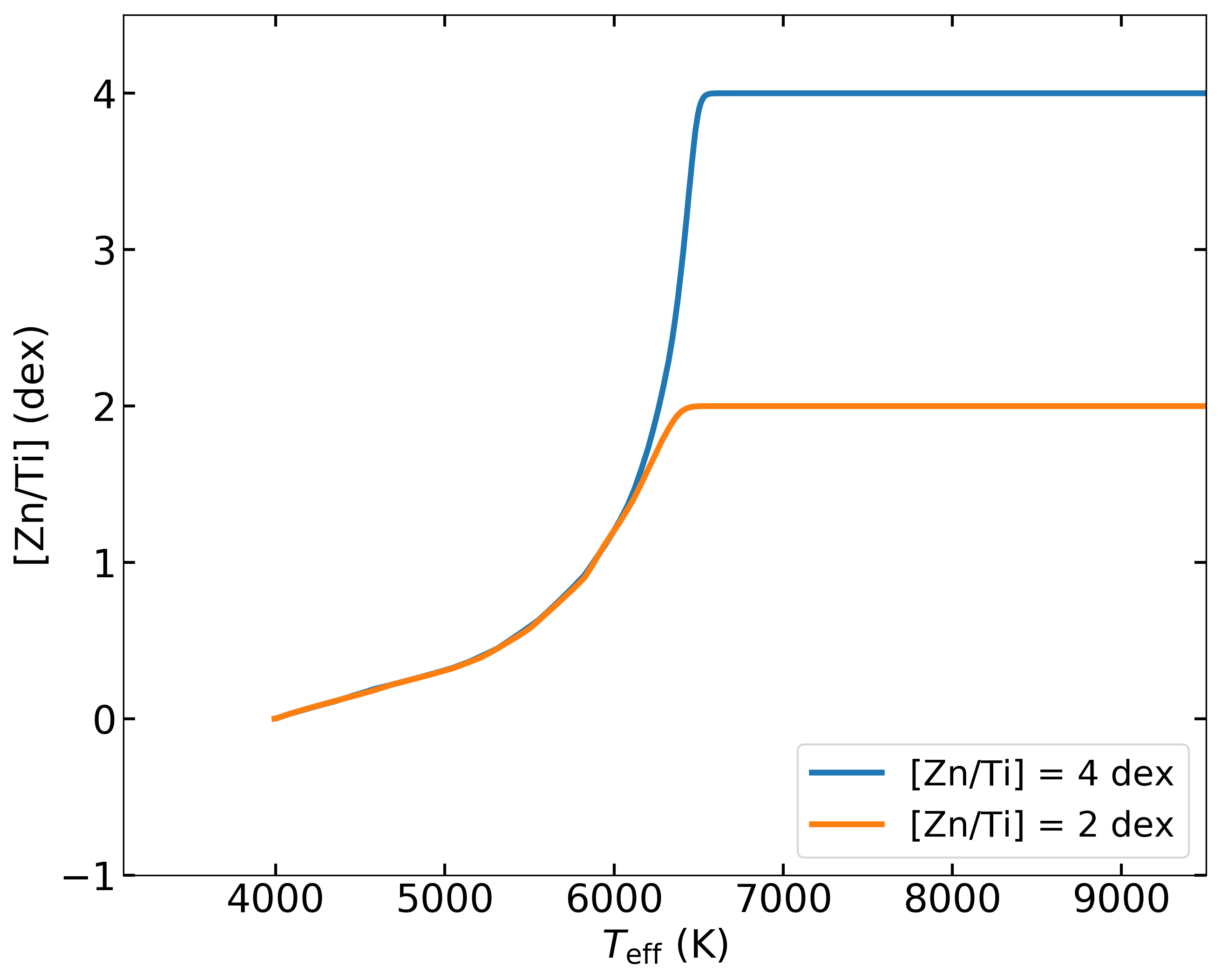}}
\caption{Two models of a 0.55-$M_\sun$ star with different maximum depletion abundances. The evolution is plotted with respect to effective temperature of the star and the [Zn/Ti] ratio in the photosphere of the model. The blue curve with [Zn/Ti] $=$ 4~dex is the accretion abundance from Fig.~\ref{fig:chemabun} which is used throughout all of our \texttt{MESA} models. The orange curve is the same model, but the accretion abundances are scaled to [Zn/Ti] $=$ 2~dex. The accretion parameters used in both models are $\dot{M}(0) = 1.5\times10^{-7}$~$M_\sun$/yr, $M_\mathrm{d} = 10^{-2}$~$M_\sun$, and $T_0 = 4000$~K.}
\label{fig:lowZmodel}
\end{figure}

In our models, we accrete gas with the same chemical composition for all stars. Since we based the abundances used in our gas-dilution model on the most depleted post-AGB stars, the turn-off temperature of the pattern in Fig.~\ref{fig:chemabun} is 800~K. However, the range in turn-off temperatures in Table~\ref{CHEMDATA} is 800--1500~K. Consequently, using only one set of accretion abundances to model the whole sample of post-AGB stars is clearly a limitation. However, since the composition of Fig.~\ref{fig:chemabun} is based on the most depleted post-AGB stars in the sample, our models show the maximum depletion a post-AGB star can reach at any given temperature. This is illustrated in Fig.~\ref{fig:lowZmodel}, where we show a comparison between a regular model and a model with less-depleted accreted gas. We plot the models in the $T_\mathrm{eff}-$[Zn/Ti] plane, which are both observable quantities and hence will allow us to compare our model results to observed stars.

The models are very similar up to the temperature at which the less-depleted model becomes saturated, because they trace the dilution factor of the convective region in the post-AGB star. This means that the only difference in changing the composition of the accreted gas of the model will be to change the moment at which the post-AGB photosphere saturates. Consequently, an observed post-AGB star can be explained by any given model as long as it falls below the curve of such a model in the $T_\mathrm{eff}-$[Zn/Ti] plane, since any star below the curve can be `fitted' by choosing a gas composition with a lower [Zn/Ti] abundance than our standard composition of Fig.~\ref{fig:chemabun}. For that reason, the results do not necessarily show which parameters of the accretion model fit best, but instead they show which range of accretion rates and disc masses are required to reproduce the observed depletion values of all the stars in the sample. The stars that provide the strongest constraints on the accretion properties are then those with relatively high depletion values at low effective temperatures.

\section{Results} \label{sect:results}

\subsection{Timescales} \label{sect:timescales}

In order to evaluate whether stars become depleted and at which temperature this happens, we compare the evolution timescale of a star to the timescale at which the star becomes depleted. For this purpose, we can define the evolutionary timescale at a given temperature as the current effective temperature divided by the rate of change: $\tau_\mathrm{evol} = T_\mathrm{eff}/\dot{T}_\mathrm{eff}$. Similarly, we define the timescale at which the star becomes depleted as the mass in the outer convective envelope divided by the current mass accretion rate: $\tau_\mathrm{depl} = 10 \times M_\mathrm{conv}/\dot{M}_\mathrm{accr}$. Here we added a factor of ten to get the timescale at which the gas becomes depleted by 1~dex. Since we present our results in terms of effective temperature and depletion value in the $T_\mathrm{eff}$--[Zn/Ti] plane, the values of $\tau_\mathrm{evol}$ and $\tau_\mathrm{depl}$ will determine the slope of the evolution track in this plane, hence determine whether a given model can become depleted or not.

\begin{figure}
\resizebox{\hsize}{!}{\includegraphics{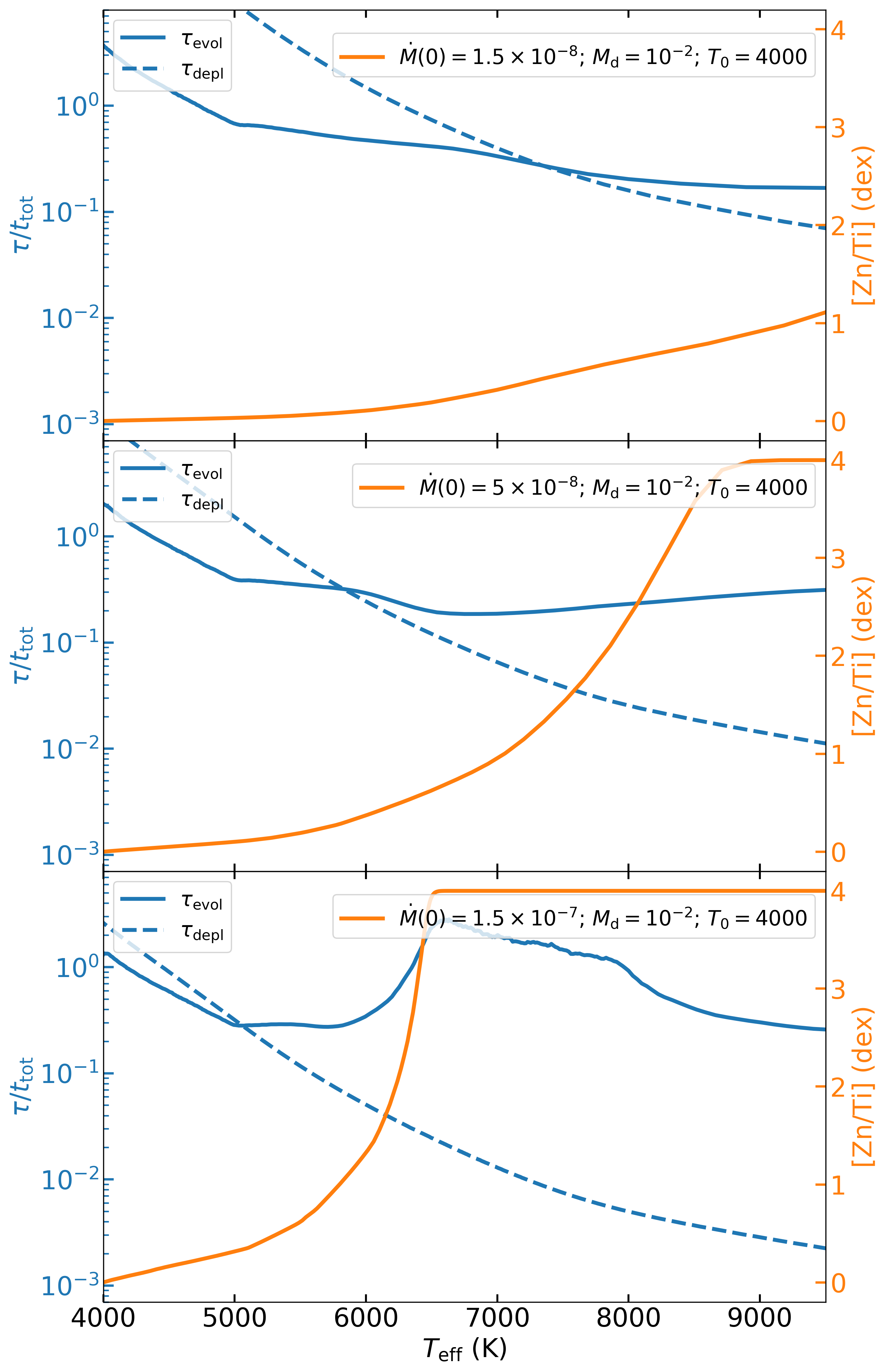}}
\caption{Depletion and evolution timescales vs depletion value for a 0.55-$M_\sun$ model for $\dot{M}(0) = 1.5\times10^{-8}$, $5\times10^{-8}$, and $1.5\times10^{-7}$~$M_\sun$/yr (top, middle, and bottom panel, respectively) with $M_\mathrm{d} = 10^{-2}$~$M_\sun$ and $T_0 = 4000$~K. Orange solid lines show the evolution of the models in the $T_\mathrm{eff}$--[Zn/Ti] plane. Blue solid and dashed lines show evolution and depletion timescales, respectively. These timescales have been normalised by dividing by the total evolution time of the model (7000, 13\,000, and 21\,000~years for top, middle, and bottom, respectively).}
\label{fig:timescales_55core}
\end{figure}

Figure~\ref{fig:timescales_55core} shows the evolution and depletion timescales for a 0.55-$M_\sun$ model for three different initial accretion rates: $1.5\times10^{-8}$, $5\times10^{-8}$, and $1.5\times10^{-7}$~$M_\sun$/yr (top, middle, and bottom panel, respectively). We start the accretion at 4000~K for all 3 models and we impose a disc mass of $10^{-2}$~$M_\sun$. The orange solid line shows the predicted depletion in the photosphere of the post-AGB star. 

For the $1.5\times10^{-8}$~$M_\sun$/yr accretion-rate model in the top panel of Fig.~\ref{fig:timescales_55core}, the depletion timescale is longer than the evolution timescale in the early stages of the evolution. This means that the model will evolve towards hotter temperatures before becoming depleted, hence its motion is horizontal in the $T_\mathrm{eff}$--[Zn/Ti] plane. As the star becomes hotter, the depletion timescale reduces considerably due to the changing envelope structure. The mass in the outer convective region that mixes with the accreted material decreases exponentially, resulting in the sharp decline of the depletion timescale in each of the models in Fig.~\ref{fig:timescales_55core}. At around 7000~K, the depletion timescale becomes shorter than the evolution timescale and the model with $\dot{M}(0) = 1.5\times10^{-8}$~$M_\sun$/yr slowly starts to become depleted.

In the $5\times10^{-8}$~$M_\sun$/yr accretion-rate model (middle panel of Fig.~\ref{fig:timescales_55core}), the depletion timescale is again larger early in the evolution of the post-AGB star, but quickly becomes smaller than the evolution timescale causing the model to reach the maximum depletion of [Zn/Ti] $=$ 4~dex. We note that the depletion timescale is defined as the timescale in which the amount of accreted material is equal to ten times the mass in the convective region, which would only amount to a dilution by a factor of ten. Consequently, since the depletion and evolution timescales are still relatively similar between 4000 and 7000~K, the model only becomes mildly to moderately depleted by the time it reaches 7000~K.

In the bottom panel of Fig.~\ref{fig:timescales_55core}, we show a model with a high accretion rate. It is clearly visible that at temperatures higher than 5000~K, the depletion timescale is much shorter than the evolution timescale such that the model quickly evolves to maximum depletion. The striking feature in the bottom panel of Fig.~\ref{fig:timescales_55core} is that the steep rise in depletion of the model coincides with an increase of the evolution timescale. This increase in $\tau_\mathrm{evol}$ in the range 6000--8000~K shows that the total evolution time of this model has been significantly extended. Initially, the evolution timescale decreases between 4000--6000~K as the star loses its envelope, accelerating the temperature evolution despite the presence of accretion. But for $T_\mathrm{eff} \gtrsim 5000$~K, the stellar wind is exponentially turned off, causing mass accretion to exceed mass loss from both wind and nuclear burning when $T_\mathrm{eff} \approx6000$~K. At this point, the envelope mass and hence $T_\mathrm{eff}$ remain almost constant, delaying the evolution until the mass-accretion rate drops below the nuclear-burning rate.

\subsection{Constraints on accretion parameters} \label{sect:accretionresults}

\begin{figure}
\resizebox{\hsize}{!}{\includegraphics{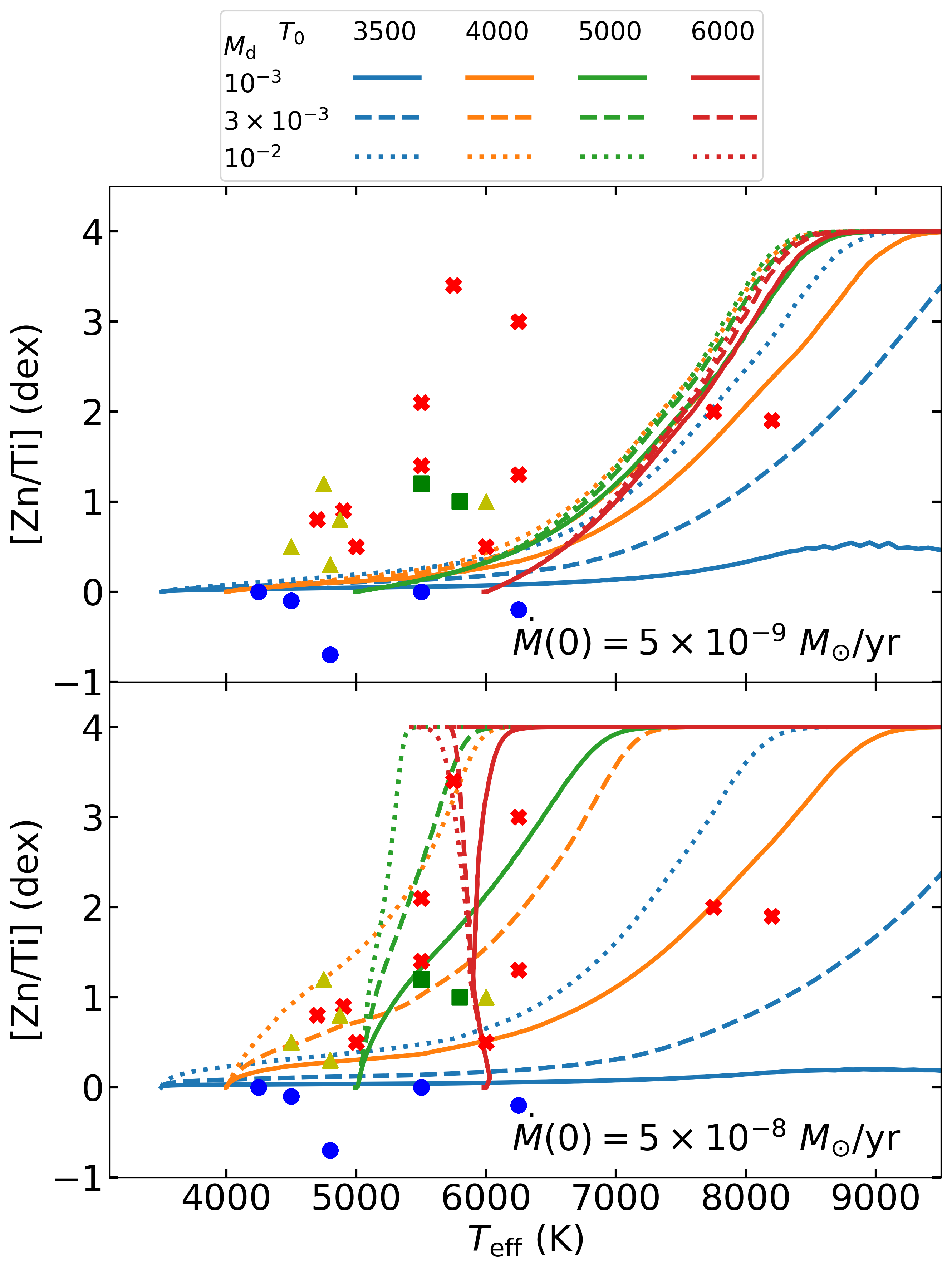}}
\caption{\texttt{MESA} models of post-RGB stars with a core mass of 0.40~$M_\sun$ and initial accretion rates of $5\times10^{-9}$~$M_\sun$/yr (upper panel) and $5\times10^{-8}$~$M_\sun$/yr (lower panel). Curves with different colours represent different starting temperatures, while linestyles distinguish different disc masses. Units of $M_\mathrm{d}$ and $T_\mathrm{0}$ are $M_\sun$ and K, respectively. Observations for stars with $L < 2500$~$L_\sun$ are shown as symbols, where similar to Fig.~\ref{fig:patterndiagram} non-depleted objects are blue circles, plateau-type objects are yellow triangles, and saturated objects are red crosses. Objects with uncertain depletion patterns are denoted by green squares. If a star does not have a Zn measurement, S is used instead, while if a star does not have a Ti measurement, we used Sc or Fe to plot the [Zn/Ti] ratio (Table~\ref{CHEMDATA}).}
\label{fig:40core}
\end{figure}

Figure~\ref{fig:40core} shows all \texttt{MESA} models with a core mass of 0.40~$M_\sun$ and initial accretion rates of $5\times10^{-9}$~$M_\sun$/yr (upper panel) and $5\times10^{-8}$~$M_\sun$/yr (lower panel). We show the models in terms of effective temperature and depletion value. The post-RGB stars that we assigned to the 0.40~$M_\sun$ bin based on their low luminosity (see Sect.~\ref{sect:massassignment}) are compared with these models of similar mass and luminosity.  

It is clear that models with low accretion rates, of the order of $5\times10^{-9}$~$M_\sun$/yr, are unable to become significantly depleted at temperatures less than 7000~K, even though many of the observed post-RGB stars show strong depletion levels at low temperatures. In fact, all our models with accretion rates lower than $5\times10^{-8}$~$M_\sun$/yr are unable to produce depleted stars at relatively low temperatures.

The models with $\dot{M}(0) = 5\times10^{-8}$~$M_\sun$/yr in the lower panel of Fig.~\ref{fig:40core} are successful in reproducing the observed depleted post-RGB stars in the sample. A striking feature is that the model with a high disc mass ($M_\mathrm{d} = 10^{-2}$~$M_\sun$) starting its evolution at 4000~K (orange dotted line) is able to become moderately depleted at effective temperatures below 5000~K. Producing moderately depleted, low-temperature stars requires both high enough accretion rates and also a slow evolution, both of which can be satisfied for post-RGB stars. 

Figure~\ref{fig:40core} reveals that the low disc-mass models are only able to become depleted quickly if they start their evolution at high temperatures. Focusing on the solid lines in the lower panel of Fig.~\ref{fig:40core}, the models starting accretion at 3500~K (blue) and 4000~K (orange) can produce only few low-depletion stars, while the 5000~K (green) and 6000~K (red) models can only reach high depletion values at around 6000~K. This suggests that the higher disc-mass models (dashed and dotted lines) are favourable for explaining the observed post-RGB stars.

\begin{figure}
\resizebox{\hsize}{!}{\includegraphics{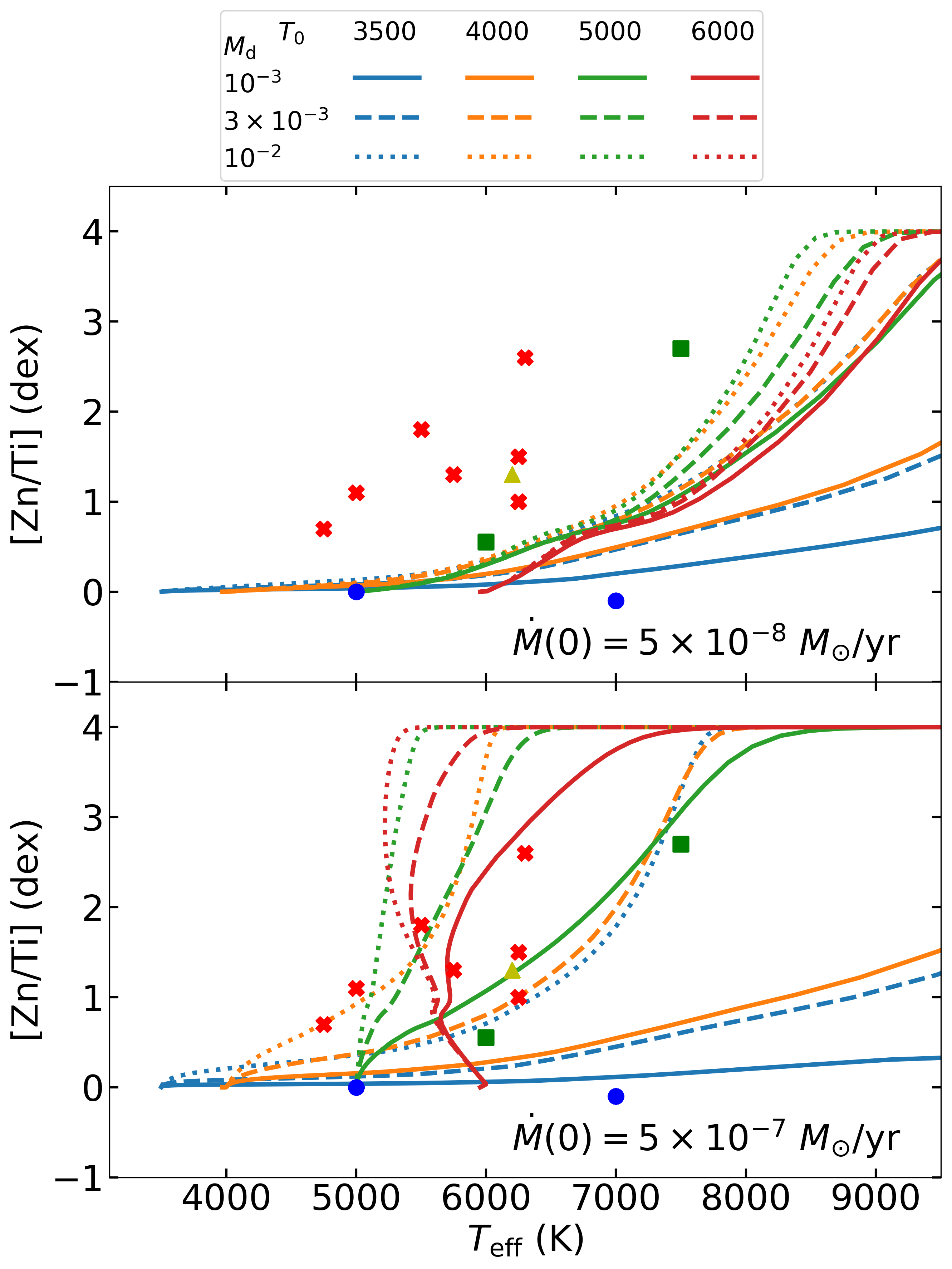}}
\caption{Same as Fig.~\ref{fig:40core}, but for post-AGB stars with a core mass of 0.55~$M_\sun$. The initial accretion rates of the models in the upper panel are $5\times10^{-8}$~$M_\sun$/yr, while in the lower panel the initial accretion rates are $5\times10^{-7}$~$M_\sun$/yr. Observed stars have a luminosity in the range 2500--7500~$L_\sun$.}
\label{fig:55core}
\end{figure}

The evolution timescale of post-AGB stars is much shorter than that for post-RGB stars. Consequently, these stars move much faster towards higher temperatures in the $T_\mathrm{eff}$--[Zn/Ti] plane, as was discussed in Sect.~\ref{sect:timescales}. Figure~\ref{fig:55core} compares the \texttt{MESA} models of 0.55~$M_\sun$ core mass to observed post-AGB stars falling in the luminosity bin in the range 2500--7500~$L_\sun$. The upper panel shows all models with an initial accretion rate of $5\times10^{-8}$~$M_\sun$/yr. This accretion rate is too low to produce the observed post-AGB systems because of the faster evolution timescales. 

Again, a higher accretion rate is required to produce the depleted post-AGB stars at low temperatures. In the bottom panel of Fig.~\ref{fig:55core}, we compare all models with $\dot{M}(0) = 5\times10^{-7}$~$M_\sun$/yr to the observed low-mass post-AGB stars. These models are successful in reproducing all the stars in that mass bin. In particular, the high disc-mass models are able to explain the mildly to moderately depleted stars at low effective temperatures. Similar to the post-RGB models, the low disc-mass models only become depleted at low $T_\mathrm{eff}$ when paired with a high starting temperature of the evolution.

\begin{figure}
\resizebox{\hsize}{!}{\includegraphics{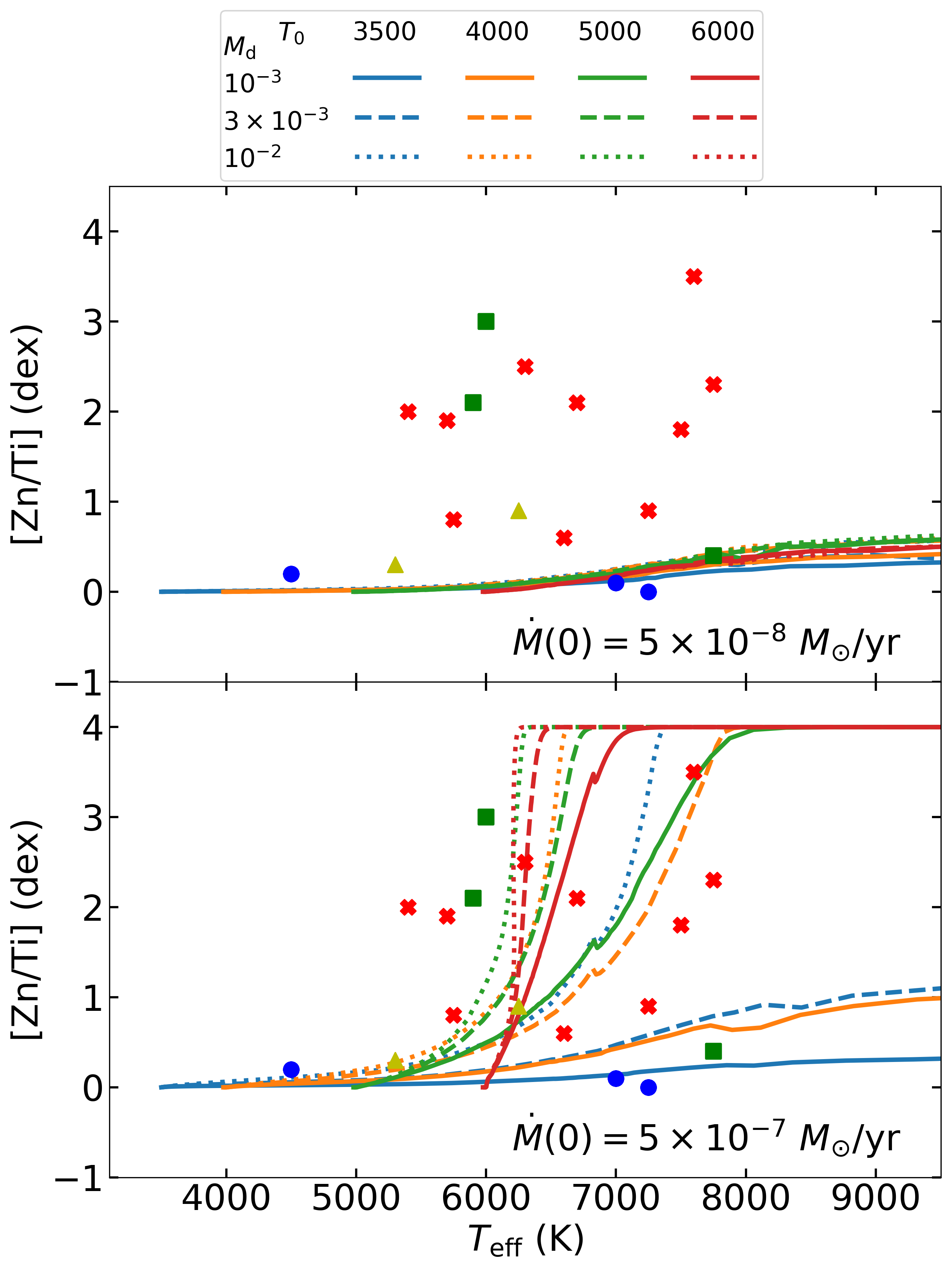}}
\caption{Same as Fig.~\ref{fig:40core}, but for post-AGB stars with a core mass of 0.65~$M_\sun$. The initial accretion rates of the models in the upper panel are $5\times10^{-8}$~$M_\sun$/yr, while in the lower panel the initial accretion rates are $5\times10^{-7}$~$M_\sun$/yr. Observed stars have a luminosity greater than 7500~$L_\sun$.}
\label{fig:65core}
\end{figure}

Finally, we compare the 0.65~$M_\sun$ \texttt{MESA} models to the highest luminosity stars in the sample. As an example, we show in the upper panel of Fig.~\ref{fig:65core} models with an initial accretion rate of $5\times10^{-8}$~$M_\sun$/yr. None of these models become depleted, regardless of the other parameters of the accretion model. Even at high effective temperatures, when the mass in the outer convective region is very small, the star still evolves too fast to become depleted. 

The bottom panel of Fig.~\ref{fig:65core} shows models with a high initial accretion rate of $5\times10^{-7}$~$M_\sun$/yr. These models are better capable of producing the observed depleted post-AGB stars. Massive post-AGB stars only seem to be able to become significantly depleted when the temperature reaches 6000~K. In these cases, the evolution timescale is prolonged similar to the bottom panel of Fig.~\ref{fig:timescales_55core}. 

Several of the luminous post-AGB stars cannot be reproduced by our models. The four depleted stars to the left of the models in the lower panel of Fig.~\ref{fig:65core} are CT~Ori, DY~Ori, HD~52961, and SS~Gem. For the RV~Tauri stars CT~Ori and DY~Ori \citep{horowitz86,manick17}, the pulsation periods indicate lower luminosities than those derived from the SED and \textit{Gaia} distance. HD~52961 and SS~Gem could be genuinely luminous stars, although the lower limit in the uncertainty of their luminosity could place them in the 0.55~$M_\sun$ mass bin. Another possibility is that our stellar wind prescription is incorrect, causing the stellar wind to fade at lower temperatures than we assume. In that case, the evolution of the star could be delayed when the star is cooler, causing the photosphere to become depleted at lower temperatures of around 5500~K.

\begin{figure}
\resizebox{\hsize}{!}{\includegraphics{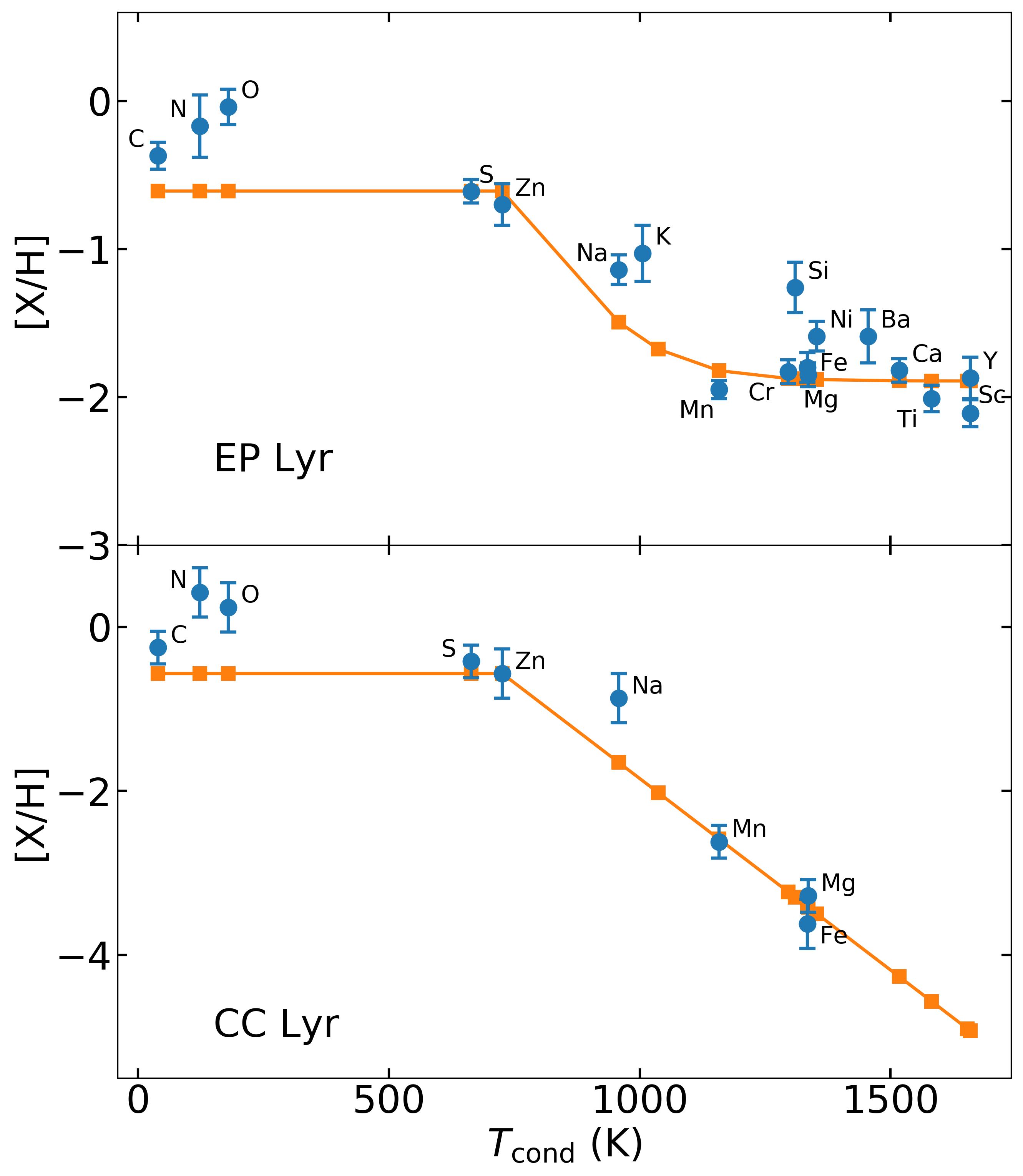}}
\caption{Comparison between the observed depletion pattern and the best-fitting model for a plateau-type profile (top, EP~Lyr) and a saturated-type profile (bottom, CC~Lyr). Observed abundances of elements are given as blue circles, while abundances from the accretion model are represented by orange squares. The elements used in the model correspond to those of Figs.~\ref{fig:chemabun} and \ref{fig:gasdil}. The model used in the top panel is a 0.55~$M_\sun$ post-AGB star starting accretion at 5000~K with $\dot{M}(0) = 5\times10^{-7}$~$M_\sun$/yr and $M_\mathrm{d} = 10^{-3}$~$M_\sun$ (Fig.~\ref{fig:55core}, green solid line in bottom panel), taken at 6200~K. In the bottom panel, the model is a 0.40~$M_\sun$ post-RGB star starting accretion at 4000~K with $\dot{M}(0) = 5\times10^{-8}$~$M_\sun$/yr and $M_\mathrm{d} = 10^{-2}$~$M_\sun$ (Fig.~\ref{fig:40core}, orange dotted line in bottom panel), taken at 6250~K. The initial metallicity of the models is normalised to the observed S and Zn abundances of both stars.}
\label{fig:eplyrcclyr}
\end{figure}

Figure~\ref{fig:eplyrcclyr} illustrates how the depletion patterns in our models compare to the observed chemical abundances of post-AGB stars. Two stars are displayed that have a similar turn-off temperature as that of our accretion model (800~K). The top panel shows EP~Lyr, which has a plateau profile, while the bottom panel shows CC~Lyr, which has a saturated profile. Based on the location of these two objects in the $T_\mathrm{eff}-$[Zn/Ti] diagrams in Figs.~\ref{fig:40core}--\ref{fig:65core}, we select the model that best fits the star. The abundances of the models shown in Fig.~\ref{fig:eplyrcclyr} are a snapshot of the chemical composition in the outer layer of the \texttt{MESA} model at the observed effective temperatures of EP~Lyr and CC~Lyr. In Fig.~\ref{fig:eplyrcclyr}, we assume that the S and Zn abundances represent the initial metallicity of the star, hence we normalise the models to these abundances.

Across all post-AGB masses, the general trend in the results is that high accretion rates are required to reproduce the observed post-AGB stars in our sample. Furthermore, low disc-mass models can only become depleted if the star starts its accretion at high effective temperatures, or in other words, in an evolved state. Since we defined the parameter of the starting temperature ($T_0$) in Sect.~\ref{sect:parameters} to account for the range of different orbital sizes we observe \citep{oomen18}, only post-AGB stars in close orbits starting their evolution at high temperatures can become depleted if the disc mass is low. However, \citet{oomen18} showed that post-AGB stars in small orbits are not depleted. These can be seen as the non-depleted objects at $T_\mathrm{eff} \gtrsim 5000$~K in Figs.~\ref{fig:40core}--\ref{fig:65core}. Consequently, since we expect post-AGB stars that start their evolution in small orbits with high $T_\mathrm{eff}$ never to become depleted in the first place, the models that are favourable for explaining the observed post-AGB sample are those with initial disc masses in the range $3\times10^{-3}$--$10^{-2}$~$M_\sun$.

These findings could provide interesting constraints on the accretion model of \citep{rafikov16} in Eq.~\ref{eq:accrrafikov}. In Sect.~\ref{sect:accretion}, we found that typical accretion rates onto the central post-AGB binary are expected to be of the order of $10^{-7}$~$M_\sun$/yr for a disc of mass $10^{-2}$~$M_\sun$. This is clearly not high enough to produce the depleted luminous post-AGB stars we observe. A higher accretion rate would require a smaller value for the initial viscous time ($t_0$) in the model. As can be seen in Eq.~\ref{eq:t0rafikov}, many physical properties enter the model. System properties of post-AGB binaries such as binary separation and binary mass are relatively well constrained \citep{oomen18}, while luminosity has little impact in determining the magnitude of the accretion rate since $\dot{M}(0)\propto L_*^{1/4}$. Instead, most uncertainty in $t_0$ comes from the parameters connected to the disc angular momentum ($\eta$ and $I_L$) and the viscosity ($\alpha$). Different assumptions for these parameters could potentially resolve the discrepancy between the expected accretion rate of the disc model and our results.

A significant result is that post-RGB stars require lower accretion rates to become depleted as compared to post-AGB stars. The reason why post-RGB stars accrete material from a disc at lower rates is unclear. In the accretion model, large differences in accretion rate are not expected between post-RGB and post-AGB stars, since luminosity is not thought to play a big role in the accretion process and the total binary mass is mainly determined by the mass of the companion. Furthermore, there is no established relation between core mass and orbital separation. Modelling the accretion rate onto post-AGB binaries likely requires a more detailed approach, where physical ingredients such as binary torque and dust formation are properly included. Finally, it is important to note that our results only impose lower limits on the disc parameters needed to reproduce the post-RGB stars, such that we cannot exclude that post-RGB stars accrete gas at similar rates as post-AGB stars.

\subsection{Evolution timescale} \label{sect:evolutiontimescale}

The luminosity of a post-AGB star greatly impacts the evolutionary timescale, as it determines the rate at which the envelope mass is reduced due to nuclear burning. Moreover, a higher luminosity generates a higher radiation pressure on the envelope, leading to stronger predicted stellar winds \citep{schroder05, cranmer11}. Because of these effects, the post-RGB/AGB evolution timescale decreases quasi-exponentially with increasing core mass.

\begin{figure}
\resizebox{\hsize}{!}{\includegraphics{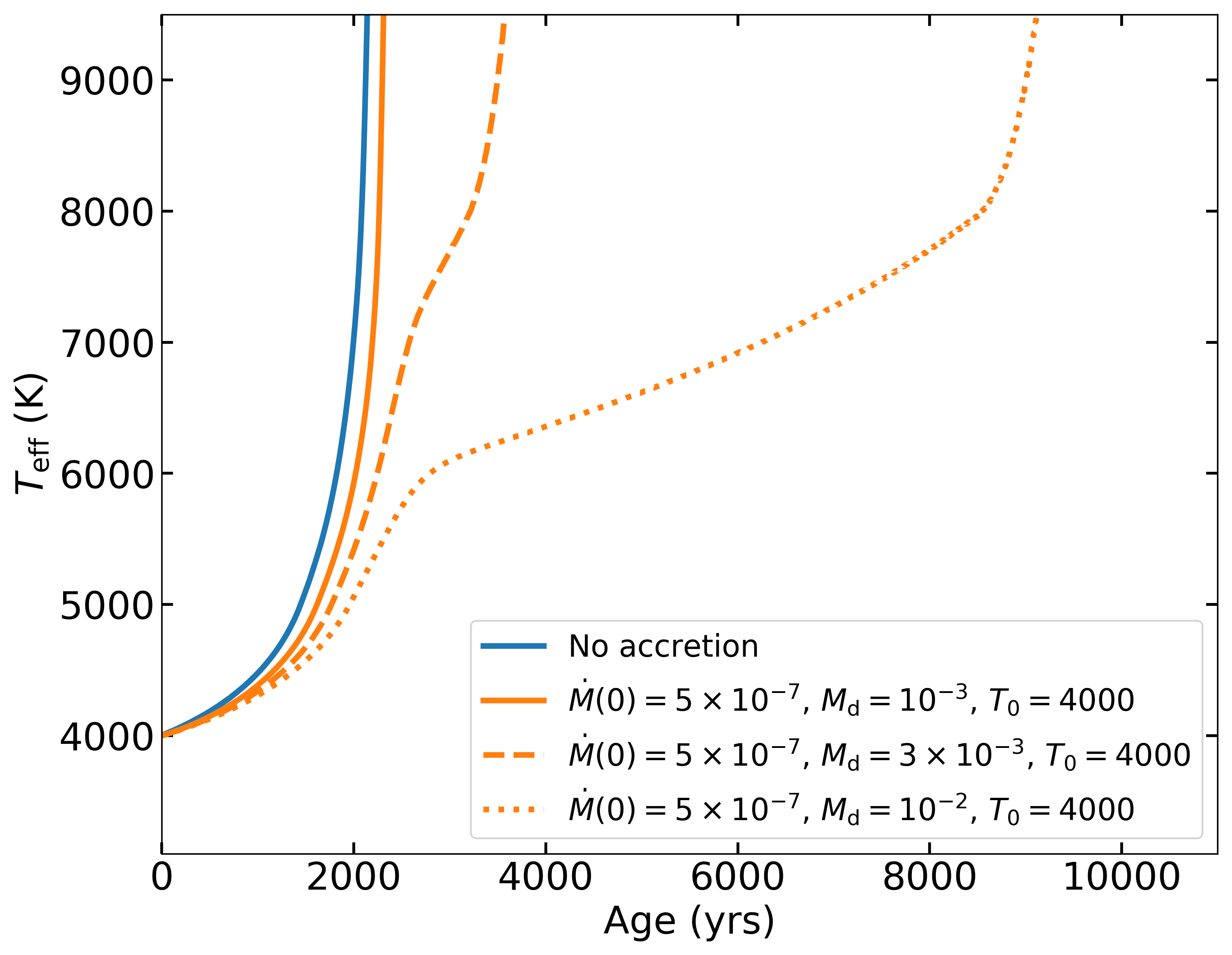}}
\caption{Evolution of effective temperature over time of a 0.60~$M_\sun$ post-AGB star for different initial disc masses and an initial accretion rate of $5\times10^{-7}$~$M_\sun$/yr. All 4 models start their evolution at an effective temperature of 4000~K. The blue solid line is a reference model without any accretion.}
\label{fig:accretiontimescales}
\end{figure}

Accreting gas onto the post-AGB star is expected to prolong the post-AGB evolution, since new hydrogen is added to the envelope while it is burned into helium at the bottom of the envelope. Figure~\ref{fig:accretiontimescales} shows the effect on the evolution timescale of a 0.60~$M_\sun$ post-AGB star for an initial accretion rate of $5\times10^{-7}$~$M_\sun$/yr and a range of initial disc masses. The figure shows that the evolution timescale of the star can be efficiently extended for high accretion rates and large disc masses. Lower disc masses have less impact on the evolution, because the accretion rate quickly drops below the nuclear burning rate.

\begin{figure}
\resizebox{\hsize}{!}{\includegraphics{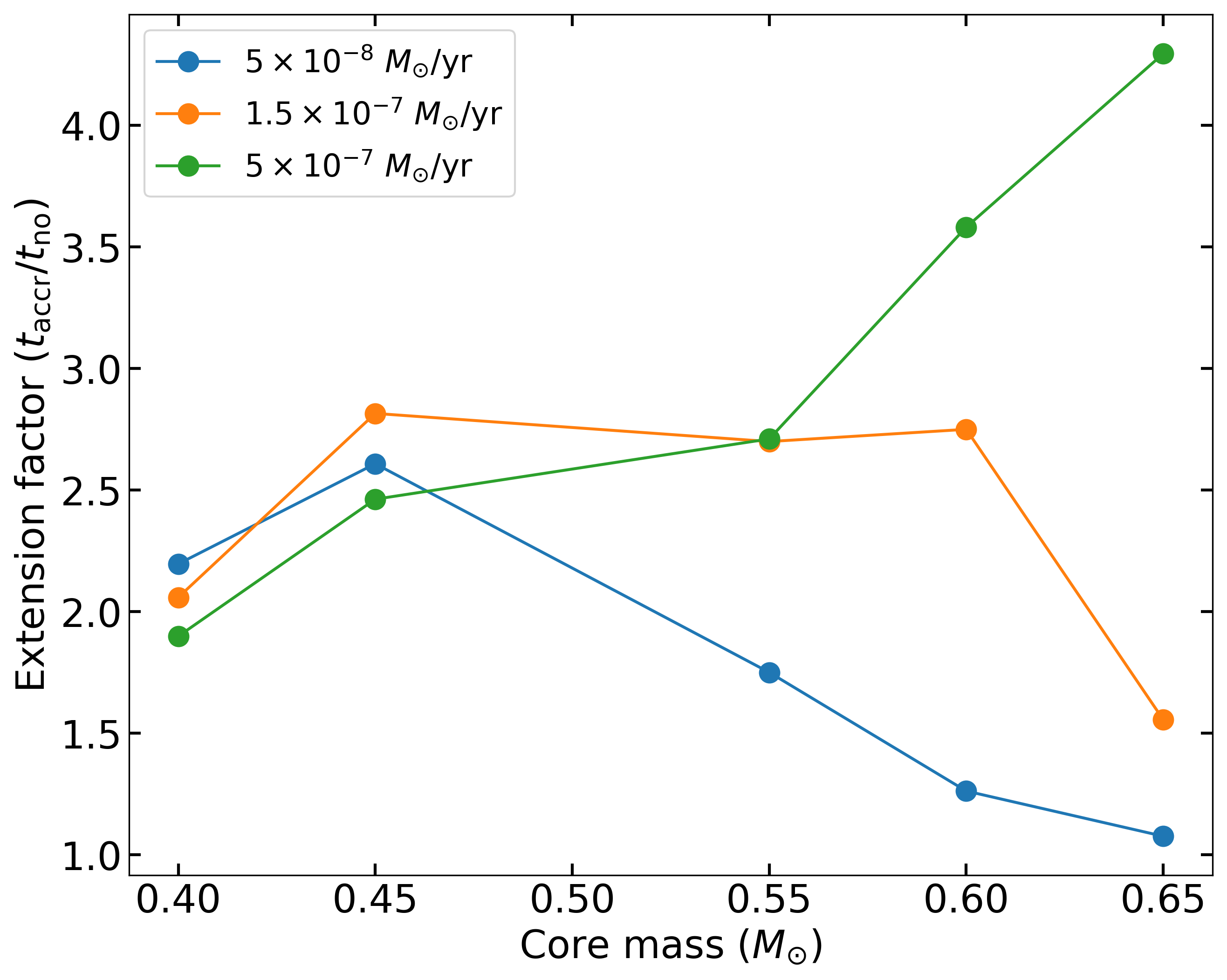}}
\caption{Factors by which the evolution of stars with different core masses are extended for three different initial accretion rates: $5\times10^{-8}$, $1.5\times10^{-7}$, and $5\times10^{-7}$~$M_\sun$/yr for blue, orange, and green, respectively. $t_\mathrm{accr}$ is the timescale at which the star evolves from $T_\mathrm{eff} = 4000$~K to $T_\mathrm{eff}=50\,000$~K while accreting gas from a disc with mass $10^{-2}$~$M_\sun$. $t_\mathrm{no}$ is the same as $t_\mathrm{accr}$ without accretion.}
\label{fig:extensionfactors}
\end{figure}

In Fig.~\ref{fig:extensionfactors}, we show by how much a large accretion rate and a large disc mass can extend the evolution for different core masses. The timescales can be extended from a factor of 2 for the post-RGB stars up to a factor of 5 for the more massive post-AGB stars. Figure~\ref{fig:extensionfactors} shows that the extension factors for post-RGB models do not depend on the initial accretion rate, since these stars evolve so slowly that they accrete similar amounts of mass within their evolution timescales, so that their lifetime is mainly determined by the initial disc mass. The more massive post-AGB models quickly move to higher temperatures, causing the stellar wind to decrease in strength and most of the mass that is accreted will need to be burned in order to decrease the envelope mass. The high accretion rate models are efficient at extending the evolution, since the accretion rate is larger than the nuclear-burning rate of the star. On the other hand, the low accretion rate models are unable to cause significant extension for the higher mass post-AGB stars, but they are also unable to become depleted (Fig.~\ref{fig:65core}, upper panel).

\subsection{Variety of depletion patterns} \label{sect:plateauvssaturation}

For all the stars in our sample (Table~\ref{CHEMDATA}), the vast majority have saturated profiles, while only a few show a plateau-like depletion pattern (middle panel of Fig.~\ref{fig:gasdil}). Some examples are EP~Lyr, HD~46703, RU~Cen, RV~Tau, and GZ~Nor (Fig.~\ref{fig:observedpatterns}, left panel). For some cases there is some ambiguity, but we identify 8 out of 58 stars to have such a plateau-like feature by comparing the underabundances of elements from the Fe group to those of the Ti group.

\begin{figure}
\resizebox{\hsize}{!}{\includegraphics{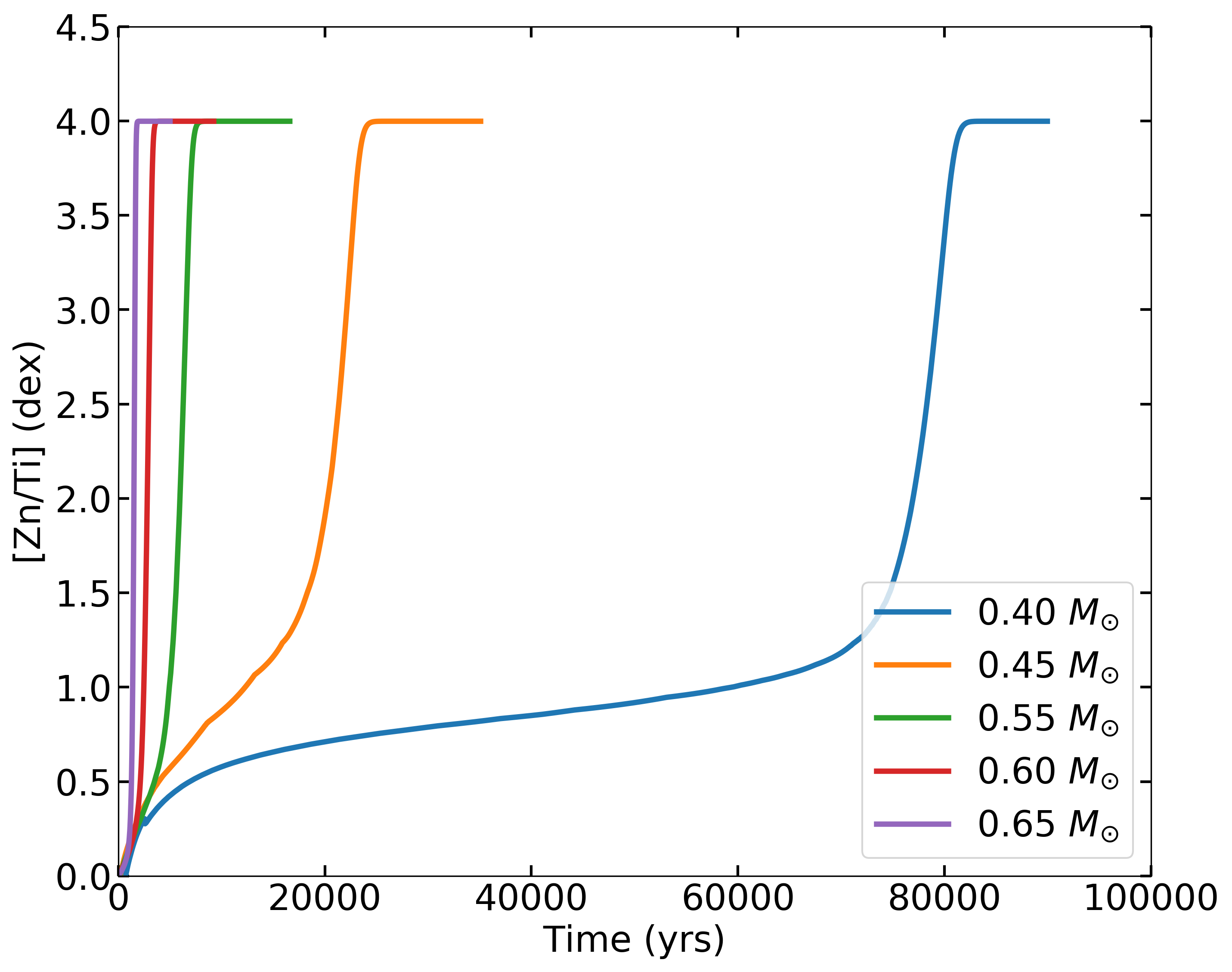}}
\caption{Time evolution of depletion for post-RGB/AGB stars of different core masses. All the models have an initial accretion rate of $5\times10^{-7}~M_\sun$/yr, an initial disc mass of $10^{-2}~M_\sun$, and start their evolution at 4000~K.}
\label{fig:abunvstime}
\end{figure}

The fact that the saturated profiles outnumber the plateau profiles can be explained by the timescale on which the atmosphere of a post-AGB star becomes depleted. In Fig.~\ref{fig:abunvstime}, we plot the abundance evolution against time instead of $T_\mathrm{eff}$. It is clear that the more massive post-AGB stars become saturated very fast, which is the reason why we expect the majority of the stars to have such a profile. Only 2 out of 18 stars classified as high-luminosity post-AGB stars show a plateau profile (AI~Sco and HD~46703), although HD~46703 is potentially misclassified since the Hipparcos parallax is much larger than the \textit{Gaia} parallax. Moreover, HD~46703 shows a small infrared excess, while we expect discs to remain massive throughout the short evolution of the high-luminosity post-AGB stars. 

Post-RGB stars, due to their slow evolution, are expected to show more plateau-like profiles. In fact, for a 0.40~$M_\sun$ core mass, the stars spend 5 times longer in their evolution with a plateau profile compared to a saturated profile (see blue curve in Fig.~\ref{fig:abunvstime}). Even though we expect a large number of plateau profiles among the post-RGB sample, we find only a small fraction (5 out of 23) of these stars to be non-saturated. However, the presence of a plateau-like depletion pattern is not only determined by the evolution timescale, but also by the composition of the accreted gas itself. In cases where the turn-off condensation temperature is high, a plateau feature cannot occur since elements such as Fe and Mg are not depleted in the first place, and only the highly refractory elements like Ti and Sc become depleted (see right panel of Fig.~\ref{fig:observedpatterns}). This can result in saturated patterns in which the chemical composition of the star is not the same as that of the accreted gas, but where instead the abundances of the highly refractory elements can still decrease as more gas gets accreted. 

Examples of this can be seen when focusing on the moderately depleted, saturated profiles of stars in the post-RGB sample in Fig.~\ref{fig:40core} with temperatures less than about 5000~K (BT~Lac, EQ~Cas, and SZ~Mon). We find that these three stars show high turn-off temperatures causing the Fe abundance to be high compared to that of Ti or Sc, which in turn leads to the classification of the star as having a saturated profile. The apparent lack of post-RGB stars with plateau-type depletion patterns could thus be caused by the high turn-off temperatures in some of these post-RGB systems. Moreover, we note that these results depend on the efficiency of mixing in stellar atmospheres (see Sect.~\ref{sect:atmosphericmixing}).

A remarkable difference between post-RGB and post-AGB stars is the presence of mildly to moderately depleted post-RGB stars at low $T_\mathrm{eff}$ in Fig.~\ref{fig:40core}, while luminous post-AGB stars can only become depleted once the temperature exceeds 5500~K (see bottom panel of Fig.~\ref{fig:65core}). Because of the short evolution timescale, a more massive post-AGB star requires much higher accretion rates in order to become depleted, since there is only a few thousand years in which the star can accrete from the disc. By contrast, due to the slower evolution of post-RGB stars, large levels of depletion can be reached at lower temperatures already, even for lower accretion rates.

\section{Discussion} \label{sect:discussion}

\subsection{Assumptions} \label{sect:assumptions}
The models we provide depend on several parameters and assumptions. Here, we discuss the validity of our assumptions and how each assumption impacts the results. 

\subsubsection{Single-star evolution} \label{sect:singlestarevolution}
We assume that binarity does not affect the evolution of the star in the post-AGB phase. This is because the post-AGB phase can be regarded as a phase of rapid contraction, in which the star moves from the size of an AGB star to the size of a white dwarf on a timescale of several thousand years \citep{vanwinckel03}. Taking this in consideration, we expect the post-AGB star to quickly shrink to sizes smaller than its Roche lobe, leaving the star to evolve as a single star.

For this purpose, we use the single-star module of \texttt{MESA}, and compute the models without including binary interactions. However, we note that for high accretion rates in low luminosity models that start accretion at high temperatures, the star can potentially increase in envelope mass, which will result in an increase in the radius of the star (e.g. red-dotted line in bottom panels of Figs.~\ref{fig:40core} and \ref{fig:55core}). Since we assume that the post-AGB evolution starts the moment the binary detaches, the star will initiate mass transfer if the radius increases beyond the initial radius. Even though in these cases the assumption of single-star evolution does not hold, this case of Roche-lobe overflow would always be stable and act as an additional mass-loss agent. This would keep the radius of the post-AGB star constant until the mass accretion rate drops to lower values causing the star to continue its evolution to hotter temperatures. Therefore, this phenomenon will not impact the results.

\subsubsection{Evolution history} \label{sect:evolutionhistory}
Our post-AGB models are prepared by evolving a 1.5~$M_\sun$ or 2.5~$M_\sun$ star up to a specific core mass and subsequently removing the envelope by increasing mass loss. These two initial masses have been arbitrarily chosen. We assume that the choice of the initial mass of the model does not impact the final properties of the post-AGB phase, since this does not depend on the evolution history but only on the current core mass and envelope mass. To test this assumption, we compared the evolution of models with a core mass of 0.55~$M_\sun$ prepared with different initial masses: 1.25, 1.5, 1.75, 2.0, and 2.5~$M_\sun$.

We find a spread in luminosity of about 10\% for these five different models. We find no relation between luminosity and the initial mass of the model. Instead, the main difference between the models originates from the phase in the thermal pulse cycle at which the star enters the post-AGB phase. Three of the five models happened to have a thermal pulse just before the start of the post-AGB phase. This means that their luminosity is still slowly increasing as the luminosity from the He-burning shell is decreasing\footnote{A higher helium-burning luminosity causes a lower hydrogen-burning luminosity, hence a lower total luminosity.}. Consequently, the three `late-thermal pulse' models have a lower luminosity as than the two models that did not undergo a recent thermal pulse, which is the main cause of difference. The small difference in luminosity of 10\% causes a change of about 35\% in evolution timescale, showing that the phase of the star in the thermal pulse cycle at the start of the post-AGB phase can have a significant impact on its properties.

To conclude, post-AGB evolution does not depend on the initial mass of the star, but on the current core and envelope mass as well as on the phase in its thermal pulse cycle at the start of the post-AGB phase. However, since we classified all observed post-AGB stars in luminosity bins which are much larger than the spread of 10\% for the different models, we argue that this does not impact our results significantly.

\subsubsection{Atmospheric mixing} \label{sect:atmosphericmixing}
Accretion is a 3-dimensional problem, since material is accreted in the equatorial plane and subsequently mixed in the atmosphere of the star. To account for this in our 1D \texttt{MESA} model, we assume that any material that is accreted mixes in regions of the star that are convective. In the early stages of post-AGB evolution, the entire outer envelope is convective, similar to an AGB star. However, as the star evolves to higher temperatures, this outer convective envelope fragments into smaller convective regions connected to sources of high opacity, such as the iron opacity peak ($\sim$200\,000~K), the second helium ionisation zone ($\sim$50\,000~K), and the \element[-]{H} opacity peak ($\sim$10\,000~K). 

\begin{figure*}
\resizebox{\hsize}{!}{\includegraphics{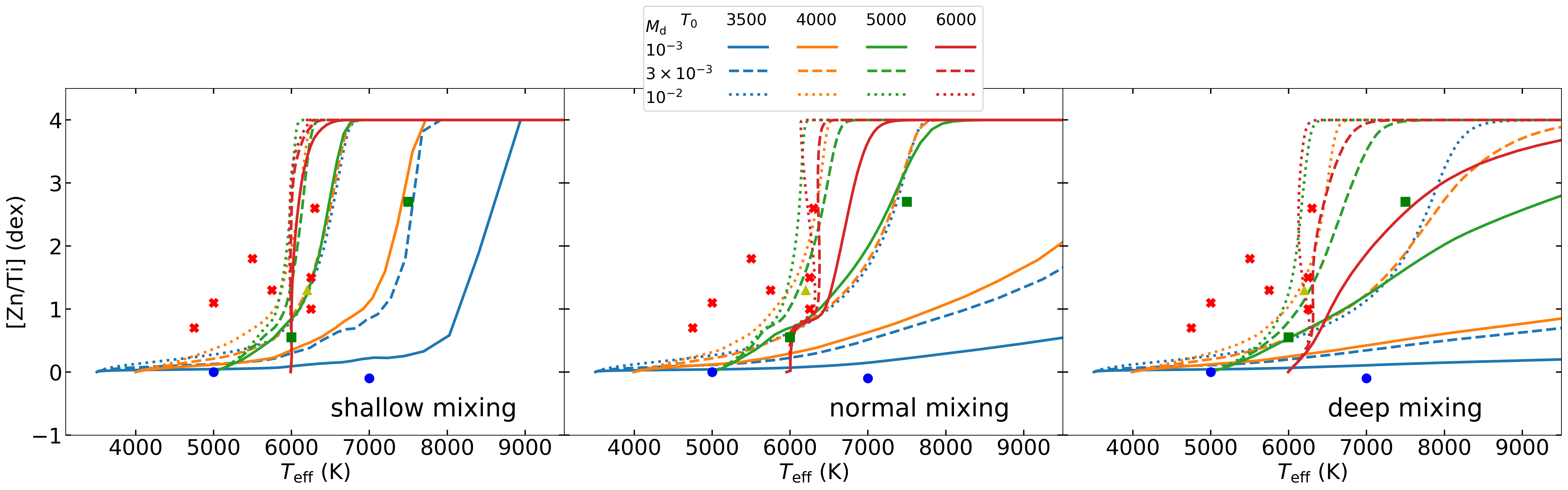}}
\caption{Comparison of 0.55-$M_\sun$ models with $1.5\times10^{-7}~M_\sun$/yr initial accretion rate for different assumptions on the depth of mixing. The left panel shows models where accreted gas is only mixed with the outermost convective zone. The middle panel shows mixing down to the \ion{He}{ii} convective zone. The right panel shows mixing down to the deep convective regions (at $T \sim 200\,000~K$).}
\label{fig:mixing}
\end{figure*}

In order to ensure a smooth evolution of the depletion patterns as matter is accreted, we decided to mix the accreted material down to the convective zone from the \ion{He}{ii} opacity peak. This means we effectively mix the outer layers of the star with temperatures less than about 80\,000~K. However, this is not necessarily a correct assumption, as it might be that the accreted gas only mixes with the outer convective zone. Alternatively, it could be that the accreted gas mixes down to deeper layers in the envelope of the star. In Fig.~\ref{fig:mixing}, we compare how the assumption of mixing down to the \element[-]{H} convective zone, the \ion{He}{ii} convective zone, or the high Fe opacity zone affects the onset of depletion in our models.

At low temperatures, the difference between the three mixing conventions shown in Fig.~\ref{fig:mixing} is negligible. This is because the envelope is still completely convective such that accreted gas is mixed down to much deeper layers in the envelope. The results start to deviate from each other once the effective temperature increases to more than 6000~K and the convective envelope starts to fragment into separate convection zones. The models with shallow mixing become maximally depleted almost instantly as the temperature increases beyond 7000~K, regardless of the accretion rate at this stage. Models with deep mixing become depleted more slowly compared to the normal models, but maximum depletion is still reached in most cases.

To conclude, the amount of mixing in the envelope does not strongly impact our results, especially at low temperatures. However, the models with shallow mixing easily become depleted at high temperatures, meaning that even for low initial accretion rates of $10^{-7}~M_\sun$/yr or lower, stars can become depleted when the temperature rises above 7000~K. However, those models cannot reproduce the depleted stars at lower effective temperatures.

\subsubsection{Stellar winds} \label{sect:stellarwinds}
\begin{figure*}
\resizebox{\hsize}{!}{\includegraphics{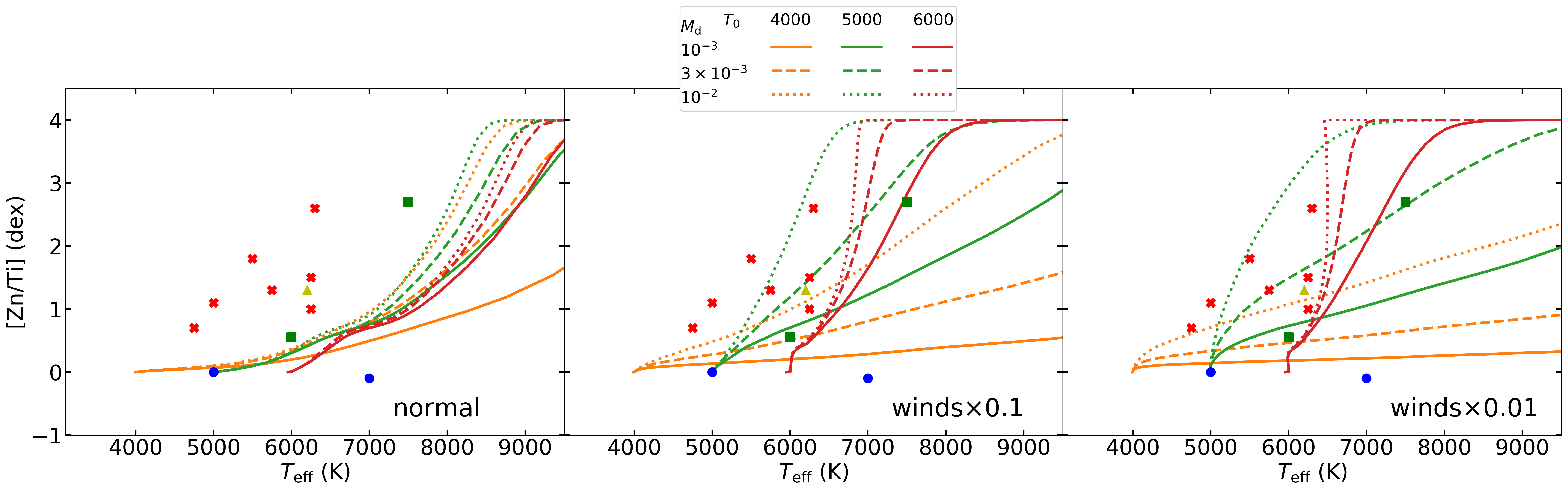}}
\caption{Comparison of 0.55-$M_\sun$ models with $5\times10^{-8}~M_\sun$/yr initial accretion rate for different stellar-wind strengths. The left panel shows models with the normal wind strength used throughout this work. The middle panel shows models with a stellar wind that is weaker by a factor of ten, while the right panel shows models with two orders of magnitude weaker wind.}
\label{fig:winds}
\end{figure*}

As was mentioned in Sect.~\ref{sect:inputphysics}, stellar winds are a major unknown factor in post-AGB evolution. In order to assess the impact of stellar winds on the results, we show in Fig.~\ref{fig:winds} models with different wind mass-loss rates. The left panel shows models with the standard wind parameters used in our results. The middle panel shows the results for winds that are weakened by a factor of 10, while the right panel shows models with a 100 times weaker wind. 

The left panel of Fig.~\ref{fig:winds} shows that for models with a typical Reimers-type wind, the evolution is too fast for the star to become depleted at an initial accretion rate of $5\times10^{-8}$~$M_\sun$/yr. However, when the wind is one or two orders of magnitude weaker, the evolution slows down enough for the stars to show depleted photospheres in the range 5000--6000~K. 

The general effect of weaker winds is to slow down evolution at temperatures less than 6000~K. Even though this effect can be significant for some cases, in the more luminous post-AGB stars ($\geq 0.60$~$M_\sun$) envelope mass loss due to nuclear burning becomes more important such that low accretion rates will not be effective at delaying the evolution (see Fig.~\ref{fig:extensionfactors}) and hence depleting these stars, regardless of the wind strength.

\subsection{Connection to orbits} \label{sect:connectiontoorbits}   

Observations show a large diversity in depletion patterns. We have already touched upon how gas dilution and mixing change the observed depletion patterns in post-AGB photospheres. Another important factor is the composition of the accreted material itself, which determines the turn-off temperature and the maximum depletion value of the pattern.

Currently, it is unknown what causes the diversity in accretion compositions. \citet{oomen18} showed that binary post-AGB stars in close orbits are not depleted. This suggests that the size of the orbit plays a role in the accretion composition, such that for the closest orbits the gas is not depleted at all. The connection to orbits is not unexpected, since the size of the orbit is expected to affect the relative distribution of gas and dust in the disc. Numerous hydrodynamics simulations of circumbinary discs show that the radius of the inner rim of the disc is located at twice the orbital separation \citep{macfadyen08,shi12,dorazio13,farris14,shi15}. On the other hand, interferometric observations of post-AGB stars are sensitive to the dust distribution in the disc, for which the inner rim is located at the dust sublimation radius \citep{hillen16}. This radius is not determined by the binary properties, but by the luminosity of the post-AGB star itself, as this determines the temperature structure by $R(T_\mathrm{dust}) \propto \sqrt{L_*}$. 

Since the inner rim of the gaseous disc depends on the binary properties while the inner rim of the dusty disc depends on the stellar properties, the interplay between these two quantities could result in the diversity of the chemical composition of the gas that eventually becomes accreted. For example, if the dust sublimation radius is larger than the inner radius of the disc, then there would be an optically thin gaseous disc inside the dusty disc, which could result in an accretion composition with a high turn-off temperature (because of the higher disc temperatures), and lower maximum depletion (because some gas will not have formed dust at all). If, on the other hand, the dust sublimation radius is smaller than the inner radius of the disc, then the entire disc would contain dust, which means that the gas itself can become more depleted, but also more elements would show underabundances since the temperature in the inner parts of the disc will generally be lower.

Essentially, we expect the diversity in accretion abundance compositions to be due to the different disc environments from which the accreted gas originates. We note that it is currently not straightforward to test this hypothesis due to several points. First of all, in order to accurately determine the luminosity, we need reliable \textit{Gaia} parallaxes. Since we are dealing with wide binaries, this will have to wait until the third data release, where the binary motion is taken into account for the parallax determination. Furthermore, only for a handful of post-AGB binaries have accurate orbital separations been determined, since these depend on both the orbital period and the masses of both components in the binary system. The most suitable systems for a quantitative analysis are those for which the inclination angle has been determined via interferometry or the RVb phenomenon \citep{manick19}. Finally, in order to qualitatively assess the accretion composition, knowledge is required on the temperature structure of the disc, which dust species form where in the disc, how much mixing occurs inside the disc, etc. Conversely, investigating depletion patterns in stars with known orbital and stellar properties could prove a valuable tool to investigate the uncertain physics in circumbinary discs.

\section{Conclusions} \label{sect:conclusions}

We have performed a comprehensive study of the formation of depletion patterns in post-AGB photospheres using state-of-the-art \texttt{MESA} models. We compared these models to a sample of 58 observed disc-type post-AGB stars with abundances from literature. We can summarise our main conclusions as follows:

\begin{itemize}
    \item We show, using detailed stellar evolution models, that the mechanism of re-accretion of metal-poor gas from a disc is capable of reproducing the observed depletion values in post-AGB stars. In order to explain the highly depleted post-AGB stars with low effective temperatures, high initial accretion rates ($\gtrsim 3\times10^{-7}$~$M_\sun$/yr) and high initial disc masses ($\sim10^{-2}$~$M_\sun$) are required.
    \item Based on these high initial accretion rates and disc masses, the post-AGB evolution timescale can be significantly extended by a factor of two for the post-RGB systems to a factor of five for the more massive post-AGB systems. Because of the large effect of gas accretion on the post-AGB timescale, it is imperative to take this effect into account when modelling the formation of planetary nebulae.
    \item We characterise the shape of an observed depletion pattern based on a simple model of gas dilution. Depleted post-AGB stars can be differentiated based on the presence of a saturated pattern or an unsaturated, plateau-type depletion pattern. We find that the diversity in patterns is not only caused by different levels of dilution of gas in the envelope of the post-AGB star, but also by different compositions of the accreted gas. The latter is probably the result of different disc structures and is possibly connected to the orbital properties of post-AGB binary systems.
    \item Differences between post-RGB and post-AGB systems are caused by large differences in luminosity and hence evolution timescale. Because of the slow evolution of post-RGB stars, we expect that these stars are much more likely to show an unsaturated, plateau profile in their abundance pattern. Although only about 20\% of observed post-RGB stars have a plateau pattern, this could be the result of large turn-off temperatures for some of these stars. We also find that post-RGB stars can become mildly to moderately depleted at relatively low effective temperatures ($\lesssim 5000$~K), while more massive post-AGB stars evolve too fast and can only become depleted at temperatures of 6000~K or larger. This is corroborated by observations of several depleted post-RGB stars at low $T_\mathrm{eff}$ and the lack of depleted post-AGB stars at $T_\mathrm{eff} < 6000$~K.
\end{itemize}

\begin{acknowledgements}
GMO acknowledges support of the Research Foundation - Flanders under contract G075916N and under grant number V434818N.
HvW acknowledges support from the Research Council of the KU Leuven under grant number C14/17/082.
This research has made use of the SIMBAD database,
operated at CDS, Strasbourg, France. This research has made use of NASA's Astrophysics Data System Bibliographic Services. We thank the anonymous referee for the useful comments that improved the manuscript.

\end{acknowledgements}

\bibliographystyle{aa}
\bibliography{references.bib}

\begin{appendix}
\section{Observational data} \label{appendix:data}
In this appendix we present the observational data collected from literature. Table~\ref{CHEMDATA} contains a list of all the stars in our sample, along with spectroscopically determined effective temperatures, as well as the abundance ratios of [Fe/H], [Zn/Ti], [S/Ti], and [Zn/Fe]. We provide the reference to these values in the final column. 

For all stars in the sample, we have determined the depletion profile based on the abundance patterns of the stars (such as the ones given in Fig.~\ref{fig:observedpatterns}). Saturated profiles are assigned by `S', while plateau profiles are assigned by `P'. In cases where the distinction is not clear, we assign the star by `U', while the stars that are not depleted at all are assigned by `N'. The classification is presented in Table~\ref{CHEMDATA}, along with the turn-off temperatures of the depletion patterns. The latter were determined by eye based on the condensation temperature of the elements that are depleted.

In Table~\ref{LUMDATA}, we show distances from \citet{bailerjones18} for all stars in our sample. Luminosities derived from fitting the SED of the stars are also shown, along with the best-fitting E(B-V) value. In deriving the $1\sigma$ confidence interval on the luminosity, we assume that the uncertainty is dominated by the uncertainty on the distance. The fourth column shows the core mass bin in which the star is placed based on its luminosity (Sect.~\ref{sect:massassignment}). However, if the ratio of the integrated flux from the infrared excess to the integrated flux from the photosphere is larger than 2, we consider the luminosity to be too uncertain to assign a mass bin to the star. This ratio is shown in the fifth column of Table~\ref{LUMDATA}.

\begin{table*}
\renewcommand{\arraystretch}{0.99}

\caption{Chemical data and depletion profiles}
\centering
\begin{tabular}{l c c c c c c c c}
\hline\hline
Star name & $T_\mathrm{eff}$ (K) & [Fe/H] & [Zn/Ti] & [S/Ti] & [Zn/Fe] & Depletion profile & $T_\mathrm{turn-off}$ (K) & Reference \\
\hline
89~Her & 6600 & -0.5 & 0.6 & 0.7 & 0.1 & S & 1500 & 1 \\ 
AC~Her & 5800 & -1.5 & 1.0 & 1.2 & 0.7 & U & 1200 & 2, 3 \\ 
AD~Aql & 6300 & -2.1 & 2.5 & 2.6 & 2.0 & S & 1000 & 3 \\ 
AI~Sco & 5300 & -0.7 & 0.3 & 0.8 & 0.1 & P & 1100 & 4 \\ 
AR~Pup & 6000 & -0.9 & / & 2.6 & / & S & 1200 & 5 \\ 
AZ~Sgr & 4750 & -1.6 & / & 1.2 & / & P & 1000 & 4 \\ 
BD+39~4926 & 7750 & -2.4 & 2.0 & 2.8 & 1.7 & S & 1000 & 6 \\ 
BD+46~442 & 6250 & -0.8 & -0.2 & 0.2 & 0.0 & N & / & 7, 8 \\ 
BT~Lac & 5000 & -0.2 & 0.5 & / & 0.1 & S & 1400 & 7 \\ 
BZ~Sct & 6250 & -0.8 & 1.3 & 1.4 & 0.9 & S & 1000 & 4 \\ 
CC~Lyr & 6250 & -3.6 & / & / & 3.0 & S & 1000 & 9 \\ 
CT~Ori & 5700 & -1.9 & 1.9 & 2.0 & 1.3 & S & 1200 & 10 \\ 
DF~Cyg & 4800 & 0.0 & -0.7 & / & -0.6 & N & / & 4 \\ 
DY~Ori & 5900 & -2.2 & 2.1 & 2.5 & 2.0 & U & 1000 & 4, 5 \\ 
EP~Lyr & 6200 & -1.8 & 1.3 & 1.4 & 1.1 & P & 800 & 5 \\ 
EQ~Cas & 4900 & -0.8 & 0.9 & 0.9 & 0.4 & S & 1500 & 4 \\ 
GK~Car & 5500 & -1.3 & 1.2 & 1.5 & 0.9 & U & 1000 & 11 \\ 
GP~Cha & 5500 & -0.6 & 0.0 & 0.3 & -0.1 & N & / & 12 \\ 
GZ~Nor & 4875 & -2.1 & 0.8 & 1.5 & 1.0 & P & 800 & 11 \\ 
HD~108015 & 7000 & -0.1 & 0.1 & 0.1 & -0.1 & N & / & 13 \\ 
HD~131356 & 6000 & 0.0 & 0.55 & 0.5 & 0.2 & U & 1300 & 13 \\ 
HD~158616 & 7250 & -0.6 & 0.0 & 0.1 & 0.2 & N & / & 13, 14 \\ 
HD~213985 & 8200 & -0.9 & / & 1.9 & / & S & 1000 & 15 \\ 
HD~44179 & 7500 & -3.3 & / & / & 2.7 & U & 1000 & 15, 16 \\ 
HD~46703 & 6250 & -1.7 & 0.9 & 1.1 & 0.8 & P & 800 & 17 \\ 
HD~52961 & 6000 & -4.5 & 3.0 & 3.4 & 3.3 & U & 1100 & 7, 18, 19 \\ 
HP~Lyr & 6300 & -1.0 & 2.6 & 3.0 & 0.6 & S & 1300 & 4 \\ 
HR~4049 & 7600 & -4.8 & / & / & 3.5 & S & 800 & 15 \\ 
IRAS~06165+3158 & 4250 & -0.9 & 0.0 & / & -0.1 & N & / & 7 \\ 
IRAS~08544-4431 & 7250 & -0.3 & 0.9 & 1.0 & 0.4 & S & 1200 & 20 \\ 
IRAS~09060-2807 & 6500 & -0.7 & 0.2 & 0.0 & 0.1 & N & / & 12 \\ 
IRAS~09144-4933 & 5750 & -0.3 & / & 1.3 & / & S & 1400 & 12 \\ 
IRAS~11472-0800 & 5750 & -2.7 & 3.4 & 3.8 & 1.8 & S & 1000 & 21 \\ 
IRAS~15469-5311 & 7500 & 0.0 & 1.8 & 2.1 & 0.3 & S & 1300 & 12 \\ 
IRAS~16230-3410 & 6250 & -0.7 & 1.0 & 1.1 & 0.3 & S & 1500 & 12 \\ 
IRAS~17038-4815 & 4750 & -1.5 & 0.7 & / & 0.3 & S & 1400 & 12 \\ 
IRAS~17233-4330 & 6250 & -1.0 & 1.4 & 1.8 & 0.7 & S & 1000 & 12 \\ 
IRAS~19125+0343 & 7750 & -0.3 & 2.3 & 2.6 & 0.4 & S & 1400 & 12 \\ 
IRAS~19135+3937 & 6000 & -1.0 & 0.5 & 0.8 & 0.0 & S & 1500 & 7 \\ 
IRAS~19157-0247 & 7750 & 0.1 & / & 0.4 & / & U & 1500 & 12 \\ 
IW~Car & 6700 & -1.1 & 2.1 & 2.5 & 1.0 & S & 1100 & 22 \\ 
LR~Sco & 5750 & -0.1 & 0.8 & 0.9 & 0.1 & S & 1400 & 4, 12 \\ 
QY~Sge & 5850 & -0.3 & 1.2 & 1.2 & 0.1 & S & 1500 & 12 \\ 
R~Sct & 4500 & -0.4 & 0.2 & / & 0.2 & N & / & 23 \\ 
R~Sge & 5000 & -0.5 & 1.1 & 1.7 & 0.3 & S & 1200 & 5 \\ 
RU~Cen & 6000 & -1.9 & 1.0 & 1.3 & 0.9 & P & 800 & 24 \\ 
RV~Tau & 4500 & -0.4 & 0.5 & / & 0.4 & P & 1100 & 23 \\ 
SAO~173329 & 7000 & -0.8 & -0.1 & 0.0 & 0.0 & N & / & 6, 13 \\ 
SS~Gem & 5400 & -0.9 & 2.0 & 1.6 & 0.9 & S & 1100 & 10 \\ 
ST~Pup & 5500 & -1.5 & 2.1 & 2.0 & 1.4 & S & 800 & 25 \\ 
SU~Gem & 5250 & -0.3 & 0.5 & 0.8 & 0.1 & S & 1400 & 7 \\ 
SX~Cen & 6250 & -1.1 & 1.5 & 1.9 & 0.6 & S & 1100 & 24 \\ 
SZ~Mon & 4700 & -0.4 & 0.8 & 1.5 & 0.0 & S & 1500 & 9 \\ 
TW~Cam & 4800 & -0.5 & 0.3 & 0.6 & 0.1 & P & 1100 & 23 \\ 
U~Mon & 5000 & -0.8 & 0.0 & 0.5 & 0.1 & N & / & 23 \\ 
UY~Ara & 5500 & -1.0 & 1.4 & 1.7 & 0.7 & S & 1000 & 23 \\ 
UY~CMa & 5500 & -1.3 & 1.8 & 2.1 & 0.7 & S & 1000 & 4 \\ 
V~Vul & 4500 & -0.4 & -0.1 & 0.7 & 0.1 & N & / & 4 \\ 
\hline
\end{tabular}
\vspace{-2.8mm}
\label{CHEMDATA} 
\tablebib{
(1) \citet{kipper11}; (2) \citet{vanwinckel98}; (3) \citet{giridhar98}; (4) \citet{giridhar05}; (5) \citet{gonzalez97a}; (6) \citet{rao12}; (7) \citet{rao14}; (8) \citet{gorlova12}; (9) \citet{maas07}; (10) \citet{gonzalez97b}; (11) \citet{gezer19}; (12) \citet{maas05}; (13) \citet{vanwinckel97}; (14) \citet{desmedt16}; (15) \citet{vanwinckel95PhD}; (16) \citet{waelkens96}; (17) \citet{hrivnak08}; (18) \citet{waelkens91}; (19) \citet{kipper13}; (20) \citet{maas03}; (21) \citet{vanwinckel12}; (22) \citet{giridhar94}; (23) \citet{giridhar00}; (24) \citet{maas02}; (25) \citet{gonzalez96}}
\end{table*}

\longtab[2]{
\renewcommand{\arraystretch}{1.25}
\begin{longtable}{l c c c c c}
\caption{Distances, luminosities, infrared luminosities, reddening, and core-mass bins\label{LUMDATA}}\\
\hline\hline
Star name &  Distance (kpc) & Luminosity ($10^3$~$L_\sun$) & Core mass ($M_\sun$) & $L_\mathrm{IR}$/$L_*$ & E(B-V)\\
\hline
\endfirsthead
\caption{continued.}\\
\hline\hline
Star name &  Distance (kpc) & Luminosity ($10^3$~$L_\sun$) & Core mass ($M_\sun$) & $L_\mathrm{IR}$/$L_*$ & E(B-V)\\
\hline
\endhead
\hline
\endfoot
89~Her & 1.4$^{+0.2}_{-0.1}$ & 9.9$^{+2.4}_{-1.8}$ & 0.65 & 0.42 & 0.03 \\ 
AC~Her & 1.23$^{+0.05}_{-0.04}$ & 2.4$^{+0.2}_{-0.2}$ & 0.40 & 0.37 & 0.22 \\ 
AD~Aql & 12.3$^{+3.9}_{-2.9}$ & 11.5$^{+8.6}_{-4.8}$ & 0.65 & 0.35 & 0.52 \\ 
AI~Sco & 11.9$^{+4.9}_{-3.4}$ & 95.7$^{+94.5}_{-47.1}$ & 0.65 & 0.63 & 0.63 \\ 
AR~Pup & 2.0$^{+1.4}_{-0.6}$ & 1.2$^{+2.2}_{-0.6}$ & / & 6.39 & 0.31 \\ 
AZ~Sgr & 5.4$^{+2.8}_{-1.6}$ & 1.4$^{+1.8}_{-0.7}$ & 0.40 & 0.0 & 0.08 \\ 
BD+39~4926 & 1.9$^{+0.2}_{-0.2}$ & 0.7$^{+0.1}_{-0.1}$ & 0.40 & 0.01 & 0.07 \\ 
BD+46~442 & 3.4$^{+1.1}_{-0.7}$ & 2.1$^{+1.5}_{-0.8}$ & 0.40 & 0.23 & 0.12 \\ 
BT~Lac & 2.8$^{+0.2}_{-0.2}$ & 0.34$^{+0.06}_{-0.05}$ & 0.40 & 0.72 & 0.26 \\ 
BZ~Sct & 6.9$^{+2.4}_{-1.6}$ & 2.4$^{+2.0}_{-1.0}$ & 0.40 & 0.77 & 0.64 \\ 
CC~Lyr & 7.4$^{+1.4}_{-1.1}$ & 0.9$^{+0.4}_{-0.2}$ & 0.40 & 0.03 & 0.08 \\ 
CT~Ori & 9.2$^{+3.0}_{-2.2}$ & 15.1$^{+11.5}_{-6.4}$ & 0.65 & 0.56 & 0.45 \\ 
DF~Cyg & 2.5$^{+0.2}_{-0.2}$ & 0.29$^{+0.05}_{-0.04}$ & 0.40 & 0.76 & 0.01 \\ 
DY~Ori & 7.7$^{+2.7}_{-1.9}$ & 21.5$^{+17.7}_{-9.5}$ & 0.65 & 0.53 & 0.95 \\ 
EP~Lyr & 5.5$^{+1.0}_{-0.8}$ & 5.5$^{+2.3}_{-1.4}$ & 0.55 & 0.05 & 0.4 \\ 
EQ~Cas & 3.8$^{+0.5}_{-0.4}$ & 0.6$^{+0.2}_{-0.1}$ & 0.40 & 0.06 & 0.21 \\ 
GK~Car & 3.8$^{+0.5}_{-0.4}$ & 1.0$^{+0.3}_{-0.2}$ & 0.40 & 0.64 & 0.41 \\ 
GP~Cha & 6.7$^{+1.2}_{-0.9}$ & 2.1$^{+0.8}_{-0.5}$ & 0.40 & 1.1 & 0.32 \\ 
GZ~Nor & 8.8$^{+3.3}_{-2.2}$ & 1.4$^{+1.2}_{-0.6}$ & 0.40 & 0.2 & 0.42 \\ 
HD108015 & 3.9$^{+0.8}_{-0.6}$ & 10.8$^{+4.9}_{-3.0}$ & 0.65 & 1.09 & 0.13 \\ 
HD131356 & 2.8$^{+0.4}_{-0.3}$ & 3.5$^{+1.0}_{-0.7}$ & 0.55 & 0.67 & 0.17 \\ 
HD158616 & 5.1$^{+1.4}_{-0.9}$ & 13.9$^{+8.8}_{-4.7}$ & 0.65 & 0.23 & 0.59 \\ 
HD213985 & 0.64$^{+0.03}_{-0.02}$ & 0.14$^{+0.01}_{-0.01}$ & 0.40 & 0.36 & 0.14 \\ 
HD44179\footnote{Since there is no parallax available for HD~44179, distance and luminosity data are taken from \citet{menshchikov02}.}  
        & 0.0$^{+0.0}_{-0.0}$ & 6.0$^{+1.2}_{-1.2}$ & 0.55 & 17.8 & 0.15 \\ 
HD46703 & 5.0$^{+0.9}_{-0.7}$ & 7.5$^{+3.1}_{-2.0}$ & 0.65 & 0.02 & 0.13 \\ 
HD52961 & 2.8$^{+0.4}_{-0.3}$ & 8.1$^{+2.3}_{-1.6}$ & 0.65 & 0.15 & 0.0 \\ 
HP~Lyr & 7.6$^{+1.4}_{-1.1}$ & 4.9$^{+2.0}_{-1.3}$ & 0.55 & 0.99 & 0.17 \\ 
HR4049 & 1.6$^{+0.5}_{-0.3}$ & 20.3$^{+14.5}_{-7.3}$ & 0.65 & 0.18 & 0.18 \\ 
IRAS06165+3158 & 2.1$^{+0.2}_{-0.2}$ & 0.49$^{+0.1}_{-0.08}$ & 0.40 & 0.36 & 0.6 \\ 
IRAS08544-4431 & 1.5$^{+0.2}_{-0.2}$ & 13.7$^{+3.8}_{-2.7}$ & 0.65 & 0.6 & 1.3 \\ 
IRAS09060-2807 & 3.5$^{+0.4}_{-0.3}$ & 0.38$^{+0.09}_{-0.07}$ & / & 3.47 & 0.17 \\ 
IRAS09144-4933 & 2.9$^{+0.5}_{-0.4}$ & 4.2$^{+1.7}_{-1.1}$ & 0.55 & 0.67 & 1.88 \\ 
IRAS11472-0800 & 2.6$^{+0.7}_{-0.5}$ & 0.28$^{+0.17}_{-0.1}$ & 0.40 & 1.89 & 0.12 \\ 
IRAS15469-5311 & 3.2$^{+0.6}_{-0.4}$ & 17.1$^{+7.2}_{-4.5}$ & 0.65 & 0.68 & 1.36 \\ 
IRAS16230-3410 & 4.4$^{+1.7}_{-1.0}$ & 3.5$^{+3.2}_{-1.4}$ & 0.55 & 0.42 & 0.83 \\ 
IRAS17038-4815 & 4.3$^{+1.3}_{-0.8}$ & 4.8$^{+3.2}_{-1.6}$ & 0.55 & 0.85 & 0.74 \\ 
IRAS17233-4330 & 6.1$^{+3.1}_{-1.9}$ & 1.1$^{+1.4}_{-0.6}$ & / & 5.09 & 0.46 \\ 
IRAS19125+0343 & 4.1$^{+0.9}_{-0.6}$ & 11.1$^{+5.4}_{-3.2}$ & 0.65 & 1.19 & 0.83 \\ 
IRAS19135+3937 & 3.8$^{+0.4}_{-0.3}$ & 2.1$^{+0.5}_{-0.4}$ & 0.40 & 0.29 & 0.28 \\ 
IRAS19157-0247 & 6.0$^{+2.4}_{-1.5}$ & 10.3$^{+9.6}_{-4.4}$ & 0.65 & 0.69 & 0.72 \\ 
IW~Car & 1.81$^{+0.11}_{-0.1}$ & 9.2$^{+1.1}_{-0.9}$ & 0.65 & 0.69 & 0.72 \\ 
LR~Sco & 7.3$^{+3.0}_{-1.9}$ & 15.8$^{+15.7}_{-7.3}$ & 0.65 & 0.91 & 0.61 \\ 
QY~Sge & 6.1$^{+3.0}_{-2.0}$ & 1.7$^{+2.0}_{-0.9}$ & / & 8.29 & 0.68 \\ 
R~Sct & 1.3$^{+1.1}_{-0.4}$ & 22.8$^{+55.0}_{-12.3}$ & 0.65 & 0.06 & 0.16 \\ 
R~Sge & 2.3$^{+0.3}_{-0.2}$ & 3.4$^{+0.9}_{-0.6}$ & 0.55 & 0.32 & 0.35 \\ 
RU~Cen & 1.8$^{+0.2}_{-0.2}$ & 1.1$^{+0.2}_{-0.2}$ & 0.40 & 0.39 & 0.18 \\ 
RV~Tau & 1.4$^{+0.1}_{-0.1}$ & 2.0$^{+0.4}_{-0.3}$ & 0.40 & 0.48 & 0.7 \\ 
SAO173329 & 7.0$^{+1.6}_{-1.2}$ & 4.0$^{+2.1}_{-1.2}$ & 0.55 & 0.47 & 0.21 \\ 
SS~Gem & 3.1$^{+0.6}_{-0.4}$ & 8.3$^{+3.6}_{-2.2}$ & 0.65 & 0.02 & 0.48 \\ 
ST~Pup & 2.5$^{+0.2}_{-0.2}$ & 0.69$^{+0.1}_{-0.08}$ & 0.40 & 0.75 & 0.02 \\ 
SU~Gem & 2.8$^{+0.9}_{-0.6}$ & 0.6$^{+0.5}_{-0.2}$ & / & 3.1 & 0.32 \\ 
SX~Cen & 3.9$^{+0.7}_{-0.5}$ & 3.3$^{+1.3}_{-0.9}$ & 0.55 & 0.45 & 0.11 \\ 
SZ~Mon & 5.1$^{+1.8}_{-1.2}$ & 2.4$^{+2.0}_{-1.0}$ & 0.40 & 0.37 & 0.01 \\ 
TW~Cam & 1.7$^{+0.1}_{-0.1}$ & 1.4$^{+0.2}_{-0.2}$ & 0.40 & 0.49 & 0.45 \\ 
U~Mon & 1.07$^{+0.12}_{-0.1}$ & 3.3$^{+0.8}_{-0.6}$ & 0.55 & 0.57 & 0.07 \\ 
UY~Ara & 4.2$^{+0.8}_{-0.6}$ & 0.6$^{+0.2}_{-0.2}$ & 0.40 & 0.63 & 0.07 \\ 
UY~CMa & 6.8$^{+1.3}_{-1.0}$ & 3.5$^{+1.5}_{-1.0}$ & 0.55 & 0.71 & 0.29 \\ 
V~Vul & 1.7$^{+0.1}_{-0.1}$ & 2.2$^{+0.4}_{-0.3}$ & 0.40 & 0.36 & 0.32 \\ 

\end{longtable}
} 
 
\end{appendix}
\end{document}